\documentclass[a4paper,aps,prd,showpacs,twocolumn,nofootinbib,10pt]{revtex4-2}

\newcommand \sst {\scriptscriptstyle}
 
\usepackage[utf8]{inputenc}
 \usepackage[english]{babel} 
 
\usepackage[unicode,bookmarks=false,colorlinks=true,linkcolor=blue,urlcolor =blue, citecolor=blue, anchorcolor=blue]{hyperref} 
            
\usepackage{amsmath,amssymb}
\usepackage{mathrsfs}
 
\usepackage{graphicx}

\usepackage{color}

\usepackage{geometry}
\begin{document}

\title{Stability of regular black holes and other compact objects with a charged de Sitter core and a surface matter layer}

\author{Angel D. D. Masa\footnote{angel.masa@ufabc.edu.br}, 
Enesson S. de Oliveira\footnote{enesson.oliveira@ufabc.edu.br},
Vilson T. Zanchin\footnote{zanchin@ufabc.edu.br}}

\affiliation{Centro de Ci\^{e}ncias Naturais e Humanas, 
Universidade Federal do ABC,\\
Avenida dos Estados 5001, 09210-580 -- Santo Andr\'{e}, S\~{a}o Paulo, 
Brazil}

\begin{abstract}
The stability and other physical properties of a class of regular black holes, quasiblack holes, and other electrically charged compact objects are investigated in the present work. 
The compact objects are obtained by solving the Einstein-Maxwell system of equations assuming spherical symmetry in a static spacetime. The spacetime is split in two regions by a spherical surface of coordinate radius $a$. The interior region contains a nonisotropic charged fluid with a de Sitter type equation of state, $p_r = -\rho_m$, $p_r$ and $\rho_m$ being respectively the radial pressure and the energy density of the fluid. The charge distribution is chosen as a well behaved power-law function. The exterior region is the electrovacuum Reissner-Nordstr\"om metric, which is joined to the interior metric through a spherical thin shell (a thin matter layer) placed at the radius $a$. The matter of the shell is assumed to be a perfect fluid satisfying a linear barotropic equation of state, ${\cal P}=\omega\sigma$, with ${\cal P}$ and $\sigma$ being respectively the pressure and energy density of the shell, with $\omega$ being a constant. The exact solutions obtained are analyzed in some detail by exploring the interesting regions of parameter space, complementing the analysis of previous works on similar models. This is the first important contribution of the present study. The stability of the solutions are then investigated considering perturbations around the equilibrium position of the shell. This is the second and the most important contribution of this work. 
We find that there are stable objects in relatively large regions of the parameter space. In particular, there are stable regular black holes for all values of the parameter $\omega$ of interest. Other stable ultracompact objects as quasiblack holes, gravastars, and even overcharged stars are allowed in certain regions of the parameter space. 


\end{abstract}

\maketitle

\section{Introduction }\label{Intro}

Black holes have been a subject of wide interest in the literature along the years. Initially, these objects attracted attention mainly because of the intriguing properties that took long to be unveiled. Since the early 1970s, the discovery by Hawking\cite{Hawking:1974sw} of quantum effects that take place near black holes and that connect gravity to thermodynamics, confirming the conjecture by Bekenstein \cite{Bekenstein:1973ur}, motivated a lot of effort to try to unveil their intriguing classical and quantum properties in full.  
 In the past few decades the interest on black holes have become even wider  since they are also solutions to several generalized gravity and grand-unifying theories, what has even turned them into conceptual objects beyond their initial realm. That is to say, nowadays the concept of black hole is found in several other theories besides the original theory of general relativity. Most of this interest has been motivated by their odd physical properties unveiled from the theoretical point of view, but also from the potential for observations of general relativistic effects related to astrophysical black hole candidates. 

More recently, the first detection of gravitational waves by the Ligo-Virgo experiment~\cite{Ligo2016} gathered much interest and attention from the community, now in view of the possibility of verifying several astrophysical aspects of black holes and other compact objects directly from observations. The analysis of the observational data collected from such an event is fully compatible with the results from simulations of the merging of two black holes within general relativity theory. 

Moreover, the first observation of the shadow of the supermassive compact object at the center of M87 galaxy, accomplished by the Event Horizon Telescope (EHT) \cite{Akiyama:2019cqa} collaboration, brought even more interest for astrophysical tests involving black holes. Once the shape of the shadows depends on the parameters of the compact object models and, moreover, the observational data is precise enough, it is possible to use such data to test and possibly ruling out some theoretical models. In the case of the M87 central object, the data analysis is compatible with the shadow of a Kerr black hole solution of general relativity.
Interestingly for the theoretical work, the analysis of the EHT data does not exclude other options of ultracompact objects that mimic black holes and, in particular, variants of the Kerr black hole that bear other parameters such as electric charge and cosmological constant, or even regular black holes \cite{Akiyama:2019fyp} are not excluded. The effects of the electric charge, cosmological constant, and the Newman-Unti-Tamburino (NUT) charge have been investigated in \cite{Grenzebach:2014fha}, while the possibility of being a regular black hole has been considered, for instance, in Refs.~\cite{Li:2013jra,Abdujabbarov:2016hnw,Dymnikova:2019vuz}. These and the future new data on strong gravitational lensing may also help us to distinguish black holes from its mimickers, including naked singularities~\cite{Virbhadra:2002ju,Virbhadra:2007kw}, see also \cite{Zulianello:2020cmx} for more references and proposed tests on black hole mimickers.

Following the perspective of the possible advances in astrophysical observations, confirmed in part by the recent developments mentioned above, and also envisaging future experiments, several possible tests to constraint the physical parameters of compact objects have been proposed during the years. Owing the purpose of the present work, the proposal by Zakharov et al.~\cite{Zakharov2005,Zakharov2005b} to extract information from supermassive black holes can be mentioned here. In particular, a procedure to constrain the electric charge parameter with current and future observations of bright stars at the Galactic Center is outlined in Ref.~\cite{Zakharov:2018awx}. The work also sets some bounds on such a parameter. However, much more stringent bounds on the electric charge of the Galactic Center black hole are found in Refs.~\cite{Zajacek:2018ycb,Zajacek:2018vsj}.

Under real astrophysical conditions, an accretion disk is formed around the compact object and then more information on the parameters of such an object may be obtained from the equilibrium conditions of the surrounding matter.
In this context, the possibility of black holes to carry some electric charge has been recently tested in simulations of the dynamics of accretion disks with electrically charged plasma \cite{Trova:2018bsf,Schroven:2018agz}. Besides, these conditions allows us to determine additional physical properties of the surrounding matter, such as the effective equation of state to model the matter of the disk (see e.g. Ref~\cite{Bronnikov2011}).
 
From the theoretical point of view, black holes are closely related to the concept of spacetime singularities. 
The inevitability of singularities, under certain physical conditions, in general relativity is a consequence of the singularity theorems~\cite{Penrose1965,HawkingPenrose1970,HawkingEllis1973,Penrose1978}. Besides formation of a singularity inside black holes, as predicted by such theorems, another important example is the big-bang singularity. There are, however, ways to avoid singularities. For instance, the quantum arguments given by Sakharov~\cite{sakharov} and Gliner~\cite{gliner1966} suggest that matter at very high densities may undergo a phase transition leading to a de Sitter phase, i.e., a phase characterized by a false vacuum where the matter pressure is negative and equals the energy density, $p =-\rho$, with the Big-Bang cosmological singularity being replaced by an initial de Sitter spacetime. This equation of state violates the strong energy condition, and then the singularity theorems do not apply. Based on this idea, several models of regular or nonsingular black holes have been proposed, see e.g.,~\cite{dy92,dy04,dy00,dy03,Bronnikov:2003yi,dy2011} and, for more references, see also Ref.~\cite{Lemos:2011vz}. Extensions of the de Sitter equation of state, especially the ones by using the inflaton field and others scalar field models, motivated many different studies on black objects free of singularities, including studies on the important problem of primordial black holes, see Refs.~\cite{ Khlopov:1985jw, Lyth:2005ze, br2006, br2007-1, br2007-2, ainou2011, Dymnikova:2015yma,Cotner:2019ykd} for a small sample of such studies.
 
The fact that some kind of exotic matter could avoid singularities formation in general relativity has been investigated since the pioneering work by Bardeen \cite{bardeen1968}. The source for the Bardeen regular black hole may be interpreted as an electromagnetic field within a particular nonlinear electrodynamics model~\cite{ab00,brcritic}.  See, e.g., Refs. \cite{ab98,br01,Hayward2006,mat06,br12,fl13,macedo14,mat08,balart14,Rodrigues:2015ayd, br2017} for a small sample of regular black hole solutions built by following Bardeen's idea, see also~\cite{anso2008,Nicolini:2008aj} for reviews on the subject of regular black holes, and 
Refs.~\cite{Ali:2018boy,Nicolini:2019irw,Neves:2019ywx,Easson:2020esh} for more recent lists of references. Many of these models present a central core that approximates a de Sitter solution, see however the recent works of Refs.~\cite{Simpson:2019mud,Berry:2020ntz} for models with asymptotically empty central cores. 

Regular black holes with a central de Sitter core may also be built within effective theories that incorporate the limiting curvature hypothesis~\cite{fmm89,fmm90,fro16}. Such theories assume the existence of a fundamental length of the order of the Planck scale, $l_{p}$, which bounds all curvature invariants, i.e., $|R|<l_{p}^{-2}$, $|R_{\mu\nu}R^{\mu\nu}|<l_{p}^{-4}$, and so on. Interestingly, the finite curvature hypothesis leads to black hole solutions free of singularities whose inner region approaches the de Sitter space.

An interesting strategy to obtain regular black hole solutions, as several cases among those reported in the above cited references, is by the matching of two different smooth spacetimes through a thin transition layer (or a surface) by convenient junction conditions. This tool, derived by Israel~\cite{Israel1966}, provides a way to analyze the characteristics and dynamics of the thin layer (or surface) with matter in the context of the general relativity. Such a strategy has been widely used in the literature to build exact solutions for compact objects of several kinds. 
In particular, several exact solutions of electrically charged regular black holes with a de Sitter core and a thin shell at the boundary have been constructed in that way (see, for instance, Refs.~\cite{Boulware:1973tlq,Lake89,Shanki:2003qm,Lemos:2011vz,uyf2012,Lemos:2017vz,Masa:2018elb, Halilsoy:2018kgy}). 

In our previous work \cite{Masa:2018elb}, we obtained regular black hole exact solutions by matching an interior de Sitter type region to an exterior Reissner-Nordstr\"om (RN) spacetime through a timelike thin shell of matter. The matter in the shell obeys a linear barotropic equation of  state, $\mathcal{P}=\omega\sigma$, $\mathcal{P}$ and $\sigma $ being the intrinsic pressure (or tension) and energy density of the shell, respectively, and $\omega$ being a constant parameter. Regular black holes and other interesting ultracompact (charged and/or uncharged) objects are found among those exact solutions. Our aim here is to examine the stability of these new regular black hole and ultracompact objects solutions having a massive thin shell. 
For instance, considering a regular black hole solution containing a thin shell of matter at the matching surface, the instability of the stationary shell immediately implies instability of the regular black hole. In this context, Balbinot and Poisson~\cite{bp90} showed that, for a certain choice of parameters, a class of uncharged regular black holes with shells~\cite{fmm89,fmm90} may be stable. Moreover, Uchikata {\it et al}.~\cite{uyf2012} applied the analysis of Balbinot and Poisson to two types of charged regular black holes: one kind with a massive but pressureless thin shell, and the other kind with a massless shell as constructed by Lemos and Zanchin~\cite{Lemos:2011vz}. They found that the black holes with a massive shell are stable solutions in a certain region of the parameter space, and in the limit of a massless shell, the configurations may also be stable against perturbations of the thin shell location. By following Ref.~\cite{bp90}, here we test the stability of the whole classes of objects found in Ref.~\cite{Masa:2018elb}.

The present work is organized as follows. In Sec.~\ref{sec:model} the basic equations of the model are implemented through the Einstein-Maxwell equations for a spherically symmetric charged fluid. The resulting system is then solved for the interior region and the solutions are briefly analyzed to complement the previous work \cite{Masa:2018elb}. The matching of the interior and exterior metrics is presented and discussed.
Section~\ref{sec:3} is devoted to analyze in detail the junction conditions and to define the matter content of the thin shell. In Sec.~\ref{sec:equilibr} we identify and describe the regions in the parameter space where regular black holes, quasiblack holes, and other interesting solutions are found. The results of the stability/instability analysis of the solutions are presented and discussed in Sec.~\ref{sec:4}. In Sec.~\ref{sec:conclusion} we make the final remarks and conclude.

Throughout this work, geometric unities such that the gravitational constant $G$ and the where speed of light $c$ are set to unity are employed, $G=1=c$, and the metric signature is $+2$.

\section{The model}\label{sec:model}

\subsection{ Basic equations and solutions}

In this paper we are mainly interested in studying the stability of the exact solutions representing regular black holes and other charged compact objects presented in Ref.~\cite{Masa:2018elb}. Aiming such a study, in this section we briefly review such solutions. 

The spacetime is considered to be static and spherically symmetric, so that the line element can be written in the form
\begin{equation}\label{eq:ds2}
ds^{2}=-B\left(r\right)dt^{2}+A\left(r\right)dr^{2}+r^{2}d\Omega^2,
\end{equation}
where  $d\Omega^2= d\theta^2 +\sin^2\theta\, d\varphi^2$ is the metric on the unit 2-sphere, $\{t,r,\theta, \varphi \}$
are Schwarzschild-like coordinates, and the potentials
$B\left(r\right)$ and $A\left(r\right)$ depend on the radial coordinate $r$ alone. 

The source is considered to be a nonisotropic charged fluid with four-velocity $U^{\mu}$ 
and preferred (anisotropy) spatial direction represented by a spacelike vector  
$X^{\mu}$. The four-vectors $U^{\mu}$ and $X^{\mu}$ 
satisfy the normalization conditions $U_{\mu}U^{\mu}=-X_{\mu}X^{\mu}=-1$, and are orthogonal to each other, $U_{\mu}X^{\mu}=0$. From these conditions and 
the metric \eqref{eq:ds2} it follows the relations
\begin{equation}
U_{\mu}=-\sqrt{B\left(r\right)}\delta_{\mu}^{t},\qquad
X_{\mu}=\sqrt{A\left(r\right)}\delta_{\mu}^{r}, 
\end{equation}
where the $\delta$ symbol stands for the Kronecker delta.

The energy density is labeled by $\rho_m$ while the radial pressure (along the direction $X_{\mu}$) and the tangential pressure (along the orthogonal directions with respect to  $X_{\mu}$) are labeled respectively by $p_r$ and $p_t$.

The electromagnetic field strength is obtained from a gauge potential which can be written as 
\begin{equation}
A_{\mu}=-\phi\left(r\right)\delta_{\mu}^{t},
\end{equation}
where $\phi\left(r\right)$ is the electric potential and depends on the radial coordinate only.

The electrically charged fluid fills the interior region, up to a limiting surface $S$, of radius $r=a$. The interior solution, for all $r<a$, is found under the assumptions
\begin{equation}\label{eq:energiaescura}
\begin{split}
p_{r}\left(r\right)+\rho_{m}\left(r\right)=0,\\ 8\pi\rho_{m}\left(r\right)+\frac{Q^{2}\left(r\right)}{r^{4}}=\frac{3}{R^{2}},
\end{split}
\end{equation}
where $R$ is an arbitrary constant parameter bearing physical dimensions of length.
The first hypothesis in Eq.~\eqref{eq:energiaescura} establishes that, in the region containing the fluid, the energy density $\rho_{m}(r)$ and the radial pressure $p_{r}(r)$ obey a de Sitter
equation of state \cite{sakharov, gliner1966}, $p_{r}\left(r\right)=- \rho_{m}\left(r\right)$,
a relation that violates some of the energy conditions.
The second hypotheses in \eqref{eq:energiaescura} establishes that, in the region containing the fluid, the 
effective energy density $\rho_{m}\left(r\right)+Q^{2}(r)/8\pi r^{4}$ 
is globally constant~\cite{coospercruz,florides83}.

An additional assumption is in respect to the charge distribution, which is in fact a necessary additional entry to close the system of equations. Following \cite{Florides1977}, the electric charge density $\rho_{e}(r)$ is chosen in the form 
\begin{equation}\label{eq:dencarga}
\rho_{e}(r)=\rho_{e0}\left(\frac{r}{a}\right)^{n}\left(1-\frac{r^{2}}{R^{2}}
\right)^{1/2},
\end{equation}
where $n\geq0$ is a dimensionless parameter and $\rho_{e0}$ is a constant carrying dimensions of electric charge per volume.

With the three above hypotheses, the system of Einstein-Maxwell equations may be solved exactly to obtain the metric potentials~\cite{Masa:2018elb}
\begin{equation}\label{eq:ABint}
A(r)=\left(1-\frac{r^{2}}{R^{2}}\right)^{-1},\quad  B(r)=\frac{1}{A(r)}, 
\end{equation}
and the fluid quantities
\begin{equation}\label{eq:denEnerIn}
8\pi\rho_{m}\left(r\right)=\frac{3}{R^{2}}-\frac{q^{2}}{a^{4}}\left(\frac{r}{a}
\right)^{2(n+1)},
\end{equation}
\begin{equation}\label{eq:prerain}
8\pi 
p_{r}\left(r\right)=-\frac{3}{R^{2}}+\frac{q^{2}}{a^{4}}\left(\frac{r}{a
}\right)^{2(n+1)},
\end{equation}
\begin{equation}\label{eq:pretanin}
8\pi p_{t}\left(r\right)=-\frac{3}{R^{2}}-\frac{q^{2}}{a^{4}}\left(\frac{r}{a
}\right)^{2(n+1)}.
\end{equation}

Besides that, the mass $\mathcal{M}(r)$ and the total electric charge inside a spherical surface of radial coordinate $r$ are, respectively,
\begin{equation}
\mathcal{M}(r)=\dfrac{r^{3}}{2R^{2}}+\dfrac{q^{2}}{2a}\left(\dfrac{r}{a
}\right)^{2n+5},
\end{equation}
\begin{equation}\label{eq:cargain}
Q(r)=q\left(\frac{r}{a}\right)^{n+3},
\end{equation}
where $q= 4\pi\rho_{e0} a^3/(n+3)$ is the total charge of the distribution. Accordingly, the electric potential is 
\begin{equation}\label{phi-int}
\phi(r)= \dfrac{q}{(n+2)a}\left[\left(\dfrac{r}{a}\right)^{n+2} +1+n\right].
\end{equation}

It is worth mentioning that the solution presented above is regular everywhere inside the matter distribution.

The region of the spacetime outside the electrically charged fluid distribution, for all $r>a$,
is electrovacuum and corresponds to a portion of the RN spacetime. Namely, the metric functions are 
\begin{equation}\label{eq:ABext}
B(r)= \dfrac{1}{A(r)}=1-\frac{2m}{r}+\frac{q^{2}}{r^{2}},
\end{equation}
where $m$ and $q$ are respectively the total mass and the total charge of the source.
The fluid quantities all vanish in this region, and the electric potential is $\phi(r) = q/r$, which matches continuously the interior solution given by Eq.~\eqref{phi-int} at $r=a$.

\subsection{ The junction conditions and the surface layer content}\label{Junction}
\label{Sec.junction}

The Birkhoff theorem allows us to join the interior de Sitter to the exterior RN spacetime regions 
by means of a dynamical (spherical) surface $\Sigma$ located at $r = a(\tau)$, where $\tau$ is a time parameter on the surface. Such a surface is considered as a thin shell that carries an uncharged perfect fluid whose energy density $\sigma=\sigma(\tau)$ and pressure (tension) $\mathcal{P}= \mathcal{P}(\tau)$ are given respectively by \cite{Masa:2018elb}
\begin{multline}\label{eq:eds}
\sigma(\tau)=-\frac{1}{4\pi 
a}\left(\sqrt{1-\frac{2m}{a}+\frac{q^{2}}{a^{2}}+\dot{a}^{2}}\right.  \\
-\left. \sqrt{1-\frac{a^
{2}}{R^{2}}+\dot{a}^{2}}\right),
\end{multline}
\begin{multline}\label{eq:pds}
\mathcal{P}(\tau) =\frac{1}{8\pi a}\left( 
\frac{a\ddot{a}-\dot{a}^{2}-1+\frac{3m}{a}-\frac{2q^{2}}{a^{2}}}{\sqrt{1-\frac
{2m}{a}+\frac{q^{2}}{a^{2}}+\dot{a}^{2}}}\right. \\
-\left. \frac{a\ddot{a}-\dot{a}^{2}-1}{\sqrt
{1-\frac{a^{2}}{R^{2}}+\dot{a}^{2}}}\right) -\sigma,
\end{multline}
where dots denote differentiation with respect to $\tau$, $\dot a= da/d\tau$, etc. 

The thin shell formalism provides
a relationship between the energy density $\sigma$ and the pressure $\mathcal{P}$ which may be written in the form $  {d\sigma}/{da}=-{2}\left(\sigma+\mathcal{P}\right)/a$ (see, e.g., \cite{Chase1970}), 
what is equivalent to the energy conservation on the thin shell.  
In fact, this relation may be cast into the form
\begin{equation} \label{eq:shell-density}
    d\left(4\pi \sigma\, a^2\right)/{d\tau}=-\mathcal{P}\,{d\left(4\pi\,a^2\right)}/{d\tau},
\end{equation}
from which we identify the total mass of the shell on the left-hand side of the equation,
\begin{equation}
M=4\pi\sigma a^2, \label{eq:shellmass}
\end{equation}
while the right-hand side may be written as $ \mathcal{P}\, d {\cal S}/d\tau$, with ${\cal S}= 4\pi\,a^2$. With this interpretation, the term on the left-hand side of Eq.~\eqref{eq:shell-density}, $dM/d\tau$, represents the variation of the internal energy, while the term on the right-hand side represents the work done by the internal forces of the shell, i.e., $dW = - \mathcal{P} d {\cal S}$.

As done in the work of Ref.~\cite{Masa:2018elb}, for the perfect fluid on the shell, we assume a barotropic equation of state of the form 
\begin{equation}\label{eq:eos}
\mathcal{P} = \omega\, \sigma,
\end{equation}
with constant $\omega$. After such a choice, Eq.~\eqref{eq:shell-density} may be integrated to yield
\begin{equation}\label{eq:EdTS}
\sigma(a)=\sigma_{0}\left(\frac{a_{0}}{a}\right)^{2\left(1+\omega\right)},
\end{equation}
where $\sigma_0$ is an integration constant satisfying the condition
$\sigma_0 = \sigma(a_0)$, with $(a_0)$ being a fixed initial position of the thin shell.

\section{Equilibrium solutions: compact objects with a massive thin shell} \label{sec:3}

\subsection{Equilibrium solutions: General properties}

Here we sort out the static case which follows by taking $a=a_0=$ constant\footnote{ Notice that, in the remaining of this section, we shall drop the ``$0$'' indexes to simplify notation.}, that means $\dot{a} =\ddot{a}=0$. In this case, Eqs.\eqref{eq:eds} and~\eqref{eq:pds} fully determine the energy-momentum content of the matching surface $\Sigma$ (a thin shell) in terms of four parameters: $a$, $R$, $m$, and $q$. If these parameters are given, the energy density $\sigma$ and the intrinsic pressure of the shell $\mathcal{P}$ result also known. However, as done in the previous work \cite{Masa:2018elb} and summarized in Sec.~\ref{Sec.junction}, we take an alternative route and impose the linear barotropic state equation~\eqref{eq:eos}. With this, a new free parameter $\omega$ is introduced in the model.

The two resulting relations from the junction conditions, Eqs.~\eqref{eq:eds} and \eqref{eq:pds}, may be used to express two out of the six fundamental parameters ($a$, $R$, $M$, $m$, $q$, $\omega$) in terms of the other four free parameters. There are, of course, a number of choices for the four free parameters, but in any case the given choice should not affect the physical interpretation of the resulting solutions. We opt to eliminate the shell mass $M$ and the total mass of the system $m$ in terms of the other four free parameters.
For this, we use relations from Eqs.~\eqref{eq:eds} and~\eqref{eq:pds}, together with Eqs.~\eqref{eq:shellmass} and \eqref{eq:eos}, and solve for $M$ and $m$ to obtain two solutions in terms of $a$, $R$, $q$, and $ \omega$. Namely,
 \begin{equation}\label{eq:massadashell4}
M_{\pm} = \dfrac{a\left(X\pm\sqrt{{X^2}^{ } - Y}\right)} {\left(1+4\omega\right)\sqrt{1-\dfrac{a^2}{R^2}}} , 
\end{equation}
and
\begin{multline}\label{eq:totalmass}
m_\pm=\dfrac{a}{2} +\dfrac{q^{2}}{2a}-
\dfrac{a}{2} \left[\sqrt{1-\dfrac{a^{2}}{R^{2}}}-\dfrac{M_\pm}{a}\right]^{2}, 
\end{multline}
where, in order to simplify notation, we introduced the quantities $X$ and $Y$, given respectively by,
\begin{eqnarray}
X& =& 2\omega\left(1-\dfrac{a^2}{R^2}\right)+\dfrac{a^2}{R^2},\\
Y& =& \left(1+4\omega\right)\left(1-\dfrac{a^2}{R^2}\right)
\left(\frac{3a^2}{R^2}-\frac{q^2}{a^2}\right).\label{eq:Y}
\end{eqnarray}

Notice that there are two independent solutions, since $M_+$ corresponds to $m_+$ and $M_-$ corresponds $m_-$, respectively, so that each pair ($M_+, m_+)$ and ($M_-, m_-)$ represents a different configuration for the same set of parameters. 

In the previous work \cite{Masa:2018elb} we performed a partial analysis of the equilibrium solutions presented above. 
In that work, we noticed that the solutions generated by $(M_-,m_-$) are interesting from the physical point of view when considering large de Sitter cores, i.e., considering configurations with $a/R$ close to unity, while the solutions generated by ($M_+,m_+$) are more interesting for small $a/R$. The interest is in solutions for which the respective masses are positive quantities.  With this in mind, for each given set of the parameters $a/R$, $q/R$, and $\omega$, 
we take here the solution given by $m =\max\, (m_-,m_+)$, and take the corresponding shell mass $M= M_{\mp}$, respectively, for each one of the choices $m=m_-$ or $m=m_+$. That is to say, we identify the largest mass between $m_\pm $ as the total gravitational corresponding to a giving set of parameters, and take $M_+$ or $M_-$ accordingly as the mass of the shell of the resulting configuration, with the solution with smaller mass being neglected.

 The functions $M_\pm$ given by Eq.~\eqref{eq:massadashell4} and, as a consequence, the functions $m_\pm$ given by Eq.~\eqref{eq:totalmass}, look as indeterminate forms in the limit $\omega\to -1/4$. In fact, such indeterminacy may be solved by substituting $\omega =-1/4$ from the beginning, in the original formulas given by Eqs.~\eqref{eq:eds} and~\eqref{eq:pds}, with $\dot a=0=\ddot a$, and Eqs.~\eqref{eq:shellmass} and \eqref{eq:eos}. Such a procedure yields \cite{Masa:2018elb}
\begin{align}
&
M={a\left(\dfrac{q^2}{a^2}-\dfrac{3a^2}{R^2}\right)}
\sqrt{1-\dfrac{a^{2}}{R^2}}{\left(1-\dfrac{3a^2}{R^2}\right)^{\!\!-1}}, \label{eq:shellmassw1o4}\\
&
m =  \dfrac{a}{2}+\dfrac{q^{2}}{2a}- \dfrac{a}{2}\left(1-\dfrac{a^2}{R^2}\right)
 \dfrac{\left(1-\dfrac{q^2}{a^2}\right)^2}
{\left(1-\dfrac{3a^2}{R^2}\right)^2  }  .
\end{align}
where the mass $m$ was obtained by replacing \eqref{eq:shellmassw1o4} into Eq.~\eqref{eq:totalmass}. These resulting expressions for $M$ and $m$ are used to analyze the case $\omega=-1/4$.

\subsection{Further conditions}\label{BHC}

An interesting feature of the present model is that the solutions for $M_\pm/R$ and $m_\pm/R$ depend explicitly on the ratios 
$a/R$ and $q^2/R^2$. Therefore,  the model presents effectively three free constant parameters, namely, $a/R\equiv a_0/R$, $q/R$, and $\omega$, with the other important parameters being given by relations~\eqref{eq:massadashell4}--\eqref{eq:Y}. 

In order to investigate the physical properties of the solutions in terms of the free parameters, a key issue is to test for the presence or absence of horizons. For instance, for a given solution to represent a regular black hole, the geometry necessarily has to present horizons. This means that quantities $r_\pm=m\pm\sqrt{m^2-q^2}$ must assume real positive values. Moreover, and more important, at least $r_+$ must be larger than the radius of the matter region boundary, i.e., $r_+/R > a/R$. Furthermore, the imposition of a timelike boundary layer (shell) implies that the condition $a/R\leq 1$ has to be imposed. Additionally, the matching of the de Sitter (inner) solution to the RN (outer) solution (see Sec.~\ref{sec:model}) has to be located inside $r_-$ and, therefore, one has the constraint $a/R\leq r_{-}/R$. In such a case, both the RN gravitational radius $r_+$ and the inner radius $r_-$ need to be in the exterior electrovacuum region, been respectively the event and Cauchy horizons. 

Even though solutions representing regular black holes are the most relevant for the present work, other configurations are also interesting. For instance, it happens that the matching may be taken arbitrarily close to the Cauchy horizon,  $a/R\to r_-/R$, giving rise to quasiblack hole configurations. Situations with no horizons as for $a/R>r_+$, corresponding to regular charged stars, and when $r_{-}$ and $r_{+}$ are not real-valued parameters, corresponding to regular overcharged stars (for which  $m^2/R^2< q^2/R^2$), are also considered in the present analysis.

\section{Analysis and classification of the equilibrium solutions}  \label{sec:equilibr}

\subsection{Preliminary remarks}\label{sec:IV-A}

As mentioned above, in the numerical analysis of the present solutions we are going to employ the normalized dimensionless variables $a/R$, $q/R$, and $\omega$.
The ranges of parameters considered in the present study are $0\leq a/R\leq 1$, $-\infty< q/R<\infty$, and $-\infty<\omega\leq 1$. The upper bound on $\omega$ is imposed by the causality condition, and negative values are allowed to consider also tension shells or thin shells made of some kind of dark fluid.  
Let us mention that, since the electric charge enters all the expressions as powers of $q^2/R^2$, without loss of generality, the numerical analysis is performed by assuming $q/R\geq 0$.

In our previous work \cite{Masa:2018elb}, the properties of the equilibrium solutions were partly investigated by means of an analysis in the $(q/R,\,a/R)$--plane. A few values of the parameter $\omega$ were selected and representative figures were drawn in each case.  
 Here, for completeness, we extend the analysis also to the $(\omega,\, a/R)$--plane by considering a few fixed values of the charge ratio $q/R$. The main results appear in Figs~\ref{f:stab-waq=0}--\ref{f:stab-waq=10}, which are representative examples. The study presented in this section is important not only to complete the previous work, but mainly to identify the important regions of the parameter space of interest for the stability analysis performed in the next section. 
 
For the sake of convenience, we separate the analysis in regions and boundaries of the regions in the parameter space.

\subsection{Boundaries in the parameter space}

\label{sec:boundaries}

\subsubsection{Preliminary remarks}

When considering the three free parameters $\omega$, $a/R$, and $q/R$, there are interesting surfaces in the parameter space that separate different regions presenting objects of different physical properties, and other surfaces that belong to the boundary of the parameter space itself. For a better visualization, we choose to show some figures in the two dimensional spaces obtained by slicing the parameter space for a few values of constant $q/R$, cf. Figs.~\ref{f:stab-waq=0}--\ref{f:stab-waq=10}, and also for a few values of constant $\omega$, cf. Figs.~\ref{f:stab-qaw=1}--\ref{f:stab-qaw=-1}.
In such two-dimensional spaces those surface boundaries appear as boundary lines.

\subsubsection{ The line $c_{\scriptscriptstyle{\mp}},\, m_-=m_+$}

This line is obtained by solving the equation $m_-(\omega,q/R,a/R) = m_+(\omega,q/R, a/R)$, cf. Eq.~\eqref{eq:totalmass}, for each given value of $q/R$ in the $(\omega,\, a/R)$--plane, and for each fixed value of $\omega$ in the $(q/R,\, a/R)$--plane. The solution is a segment of the curve given by the relation $a/R=\sqrt{1+2\omega}/\sqrt{2(1+\omega)}$, independently on the electric charge $q/R$. 
The solution is represented in all the figures (when present) by a dashed brown line, and it is also indicated by the appropriate label $c_{\scriptscriptstyle{\mp}}$. 

The curve $c_{\scriptscriptstyle{\mp}}$ extends all along the parameter space, except for $q/R=0$ where it does not appear in the region with $\omega < -1/2$ (it coincides with the line $a/R=0$). Note that the segment of  $c_{\scriptscriptstyle{\mp}}$ given by the function  $a/R=\sqrt{1+2\omega}/\sqrt{2(1+\omega)}$ is continued along the two branches of the curve $c_4$ (see Figs.\ref{f:stab-waq=0}--\ref{f:stab-qaw=-1}).

The complete line $c_{\scriptscriptstyle{\mp}}$, including the sectors where it coincides with the two branches of $c_4$,  separates the parameter space into two regions. The configurations represented by the region above such a line are obtained from the masses $m_-$ and $M_-$, while the configurations represented by the region below it are obtained from the masses $m_+$ and $M_+$. As mentioned above, this is the choice that maximizes the regions of the parameter space containing solutions representing objects with good physical properties.  Note also that this choice implies the mass of the shell $M$ is not a continuous function in the parameter space, since it presents a jump when crossing the segment of line $c_{\scriptscriptstyle{\mp}}$ given by $a/R=\sqrt{1+2\omega}/\sqrt{2(1+\omega)}$. However, the total gravitational mass $m$ is a continuous function everywhere, what guarantees the smoothness of the resulting spacetime geometries in the parameter space, even in the neighborhood of any point belonging to $c_{\scriptscriptstyle{\mp}}$.

\subsubsection{ The line $c_1,\, m/R=q/R$}

This line is the locus of extremely charged objects in the parameter space, which is obtained by substituting $m/R=q/R$ into Eq.~\eqref{eq:totalmass} and solving for $a/R$ as a function of $\omega$ for each fixed value of  $q/R$ in the $(\omega,\, a/R)$--plane, and by solving for $a/R$ as a function of $q/R$ for each fixed value of  $\omega$ in the $(q/R,\, a/R)$--plane.
The resulting equation presents real solutions just for $\omega$ in the range $-1.725\lesssim \omega\leq 1$. 
The solutions are represented by green dashed lines labeled as $c_1$ in all figures, except in Fig.~\ref{f:stab-waq=0} where it is not present.

As it can be seen from Figs.~\ref{f:stab-waq=02}--\ref{f:stab-qaw=-1}, the line $c_1$ separates the regions of undercharged from the regions of overcharged objects. We find four different instances. The line (surface) $c_1$ appears between regions $(i)$ and $(ii)$, between regions $(ii)$ and $(iii)$, between regions $(iii)$ and $(iv)$, and/or between regions $(iv)$ and $(v)$. In the first case, it bears extremely charged stars, while it bears extremely charged regular black holes in all the other three cases.

For values of charge in the interval $0< q/R < 3\sqrt{3}/4$, the solution to the resulting equation presents two branches, generating two open curves in the $(\omega,\, a/R)$--plane, see Figs.~\ref{f:stab-waq=02}--\ref{f:stab-waq=1}.
The region between the two branches of line $c_1$ and bounded by curve $c_4$ contains overcharged configurations (with $q/R > m/R$) and other less interesting solutions, while the regions above the upper branch and below the lower branch contain undercharged (with $q/R< m/R)$, more interesting solutions.

For values of the electric charge in the interval $q/R\geq 3\sqrt{3}/4$, the two branches of line $c_1$ meet each other on the line $c_{\scriptscriptstyle{\mp}}$ generating a single open curve, see Figs.~\ref{f:stab-waq=1.3}--\ref{f:stab-waq=10}. In the special case of Fig.~\ref{f:stab-waq=1.3}, for $q/R= 3\sqrt{3}/4$, the two branches of $c_1$ join each other on the boundary of the $\omega$ range, at the point $(\omega=1$, $a/R= \sqrt{3}/2)$.
The undercharged solutions are then found in the regions above and to the right of such a curve. In the limit of very large electric charge, curve $c_1$ coincides with the vertical axes $\omega=0$. For some more details see Sec.~4.3.3 of Ref.~\cite{Masa:2018elb}.

\subsubsection{ The line $c_2,\, M=0$}

This line corresponds to the class of solutions without a thin shell, i.e., for which the intrinsic mass, energy density, and pressure of the shell are all zero. In fact, Eqs.~\eqref{eq:eds} and \eqref{eq:pds} together with the conditions $M=0$ and $a/R\neq 1$ imply in $\sigma=0$ and $\mathcal{P}=0$.  In such a situation, the junction between the de Sitter interior region and the exterior RN region is made smoothly (without the thin shell), by means of a boundary surface.

The solution of the equation $M(\omega,a,q)=0$ gives $q/R= \sqrt{3}\, a^2/R^2$ for some restricted interval of values of $\omega$ that depends upon the electric charge. In all representative figures, the corresponding solutions are represented by red dashed lines labeled as $c_2$. The full real solution is a segment of the curve $q/R=\sqrt{3} a^2/R^2$, and it is well visualized in the $(q/R,\, a/R)$--plane, cf. Figs.~\ref{f:stab-qaw=1}--\ref{f:stab-qaw=-1}, where it has an extremity on the line $c_{\scriptscriptstyle{\mp}}$  and the other one at the point $(q/R=\sqrt{3},\, a/R=1)$, independently of $\omega$.
 The area bounded by this line, a segment of line $c_4$, and by the line $c_{\scriptscriptstyle{\mp}}$ contains configurations with $M/R<0$, while the remaining region of the parameter space contains configurations with $M/R>0$. For other details see Sec.~4.3.4 of Ref.~\cite{Masa:2018elb}.

Notice that, in the $(\omega,\,a/R)$--plane, the curve $c_2$ appears as a segment of the horizontal line $a/R=\left(\sqrt{3}\,q/3R\right)^{1/2}$ that starts on the line $c_{\scriptscriptstyle{\mp}}$ and ends at line $c_4$ (see below).
For electric charges in the interval $0< q/R< 1/\sqrt{3}$, the line $c_2$ lies below the curve $c_{\scriptscriptstyle{\mp}}$, while it lies above $c_{\scriptscriptstyle{\mp}}$ for $q/R$ in the interval $1/\sqrt{3}< q/R < \sqrt{3}$.
For $q/R=0$, $q/R=1/\sqrt{3}$, and in the interval $q/R> \sqrt{3}$, the line $c_2$ is not present,  see Figs.~\ref{f:stab-waq=0}--\ref{f:stab-waq=10}. 
On the other hand, the line $c_2$ is also a segment of the curve $a/R=\sqrt{q/R}/\sqrt[4]{3}$ for $\omega>-a^2/2R^2(1-a^2/R^2)$. In the $(q/R,\, a/R)$--plane, the line $c_2$ satisfies the relation $a/R=\sqrt{q/R}/\sqrt[4]{3}$, see Figs.~\ref{f:stab-qaw=1}--\ref{f:stab-qaw=-1}.

The configurations belonging to $c_2$ are similar to the particular case studied in Refs.~\cite{Lemos:2011vz,uyf2012}, whose solutions satisfy the relation $a/R=\sqrt{q/R}/\sqrt[4]{3}$ and present no thin shell, but here the electric charge is not confined to the boundary surface. 

A special case of the curve $c_2$ deserves further analysis. As depicted in Fig.~\ref{f:stab-waq=1.3}, for $q/R=3\sqrt{3}/4$, $c_2$ is the whole horizontal line $a/R=\sqrt{3}/2$ and it coincides with a branch of the curve $c_1$, implying that the solutions are extremely charged black holes without a thin shell. 

Notice also that there is another line where the relation $M/R=0$ is satisfied, this is when $a/R=1$, which is the (upper) boundary surface in the three dimensional parameter space. This fact can be verified by performing a Taylor expansion of the mass function $M/R$ around the point $a/R=1$, that gives ${M}/{R}=$ $ \left( {q^{2}}/{R^{2}}-3\right)\left(1-{a}/{R}\right)^{1/2}/ {\sqrt{2}}$ $+ \mathcal{O}\left(\left[{a}/{R}-1\right]^{3/2}\right)$ and from what follows that $M/R$ vanishes in the limit ${a/R\to 1}$.  
However, it is worth mentioning that the configurations on the line $a/R=1$ are singular due to the fact that $\mathcal{P}$ is not well defined, it diverges at all points on this line, except for the particular value $q/R=\sqrt{3}$, where $\mathcal{P}$ vanishes. For more details see Secs.~4.3.6 and 4.3.7 of Ref.~\cite{Masa:2018elb}.

\subsubsection{The line $c_3,\, r_-/R=a/R$}

This line is drawn for the condition that the boundary shell coincides with the inner radius of the RN metric, $a/R=r_-/R$. In the interval with $0<a/R<1$, the same line may be obtained by taking the condition $r_+/R=a/R$. The two conditions together imply in the relation $r_+/R=r_-/R$, which also implies the equality between the total mass and the electric charge of the solution, $m/R=q/R$, so that the solutions on this line are extreme objects.  In the $(\omega,\,a/R)$--plane, this line is a segment of the horizontal line $a/R=q/R=$ constant. In turn, in the $(q/R,\, a/R)$--plane, $c_3$ is a segment of the line $a/R=q/R$.
As seen in Figs.~\ref{f:stab-waq=0}--\ref{f:stab-waq=10}, and also in Figs.~\ref{f:stab-qaw=1}--\ref{f:stab-qaw=-1}, this line is only present for values of electric charge $0\leq q/R\leq 1$. This is a consequence of the restriction on the parameter $a/R$, which assumes values in the interval $0\leq a/R\leq 1$
 
When considering neighboring sectors of parameter space bearing undercharged solutions, with $m/R\geq q/R$, the line $c_3$ separates the region of regular charged black holes from the region of regular charged stars. This fact can be seen in all figures shown in the present section, except in the especial cases of $q/R=0$ and $q/R=1$ where it coincides with the lines $a/R =0$ and $a/R =1$, respectively,  see Figs.~\ref{f:stab-waq=0} and \ref{f:stab-waq=1}. 

For $0< a/R<1$, all physical quantities $M$, $m$, $\sigma$, and $\mathcal{P}$ are well defined on the line $c_3$, but the matching surface character depends on the observer point of view. From the external spacetime analysis, the matching is made on the extreme horizon of the RN metric, which is a lightlike surface (located at $a/R=r_-/R=m/R$), while from the inner de Sitter metric analysis, the matching surface is timelike (located at $a/R<1)$. 
According to Ref.~\cite{LemosZaslavskii2007}, the solution may be interpreted as a quasiblack hole. In fact, the matching of the two spacetime metrics would lead to $\sqrt{1-a^2/R^2}\, dT= \left(1-m/a\right) dt$, where $T$ and $t$ are respectively the interior and exterior time coordinates.  
In the limit $m/a\rightarrow 1$, the coefficient $g_{tt}=\left(1-m/a\right)^2$ vanishes while $g_{\sst TT}=1-a^2/R^2 $ does not (since we have $0 < a/R <1$), and then for any finite time interval $dT$ it elapses an arbitrarily large time interval $dt$, leading to causally disconnected spacetimes. As discussed for instance in Sec.~C 2 of Ref.~\cite{LemosZaslavskii2007} (see also \cite{LemosZaslavskii2020} for a recent short review), the whole region $0\leq r < a$ becomes an infinite redshift region and the surface $a\to m$ forms a quasihorizon, characterizing a quasiblack hole configuration. 

For $q/R=0$, the line $c_3$ coincides with the horizontal axes $a/R=0$ for all $\omega$ (see Fig.~\ref{f:stab-waq=0}) and the respective solution is the flat spacetime. For more details, see the discussion related to the boundary line $a/R=0$ given in Ref.~\cite{Masa:2018elb}.

For $q/R=1$, the line $c_3$ coincides with the boundary $a/R=1$ for all values of $\omega$ (see Fig.~\ref{f:stab-waq=1}). Here the thin shell mass vanishes, the total gravitational mass is finite, $m/R=1$, but the superficial pressure $\mathcal{P}$ diverges. Therefore, all solutions in this limit represent singular extreme quasiblack holes.

\subsubsection{The line  $c_4,\, \operatorname{Im}(M/R)=0$}

This line represents the boundary of real solutions for the thin shell mass $M/R$. In both cases, for sections of the parameter space with constant $q/R$ or constant $\omega$, it is drawn as the contour curve for zero imaginary part of $M/R$, $\operatorname{Im}(M/R)=0$. 
All the relevant quantities such as the total mass, the intrinsic energy density and pressure, and the mass of the shell are real and well defined on the curve, and then it represents interesting physical configurations. The physical properties of such objects may be inferred from the objects of the neighboring regions, whose descriptions are given in Sec.~\ref{sec:regions}.

 \subsubsection{The line $c_5,\, m/R=0$}
 This line corresponds to solutions with zero total gravitational mass, $m/R=0$, and then it represents configurations similar to regular overcharged stars. All the physical quantities are well defined on such a line. It occurs just for small negative values of $\omega$, and it appears in all figures drawn in the $(\omega,\,a/R)$--plane. In such a plane, the region between this line and the line $c_4$ contains solutions of negative total mass. The specific properties of the objects belonging to this line vary from case to case, and may be inferred from the objects of the neighboring regions, whose descriptions are given in Sec.~\ref{sec:regions}.

\subsubsection{ Other boundaries}

Besides the special boundary regions commented above,
there are some other surfaces (that appear as lines in the $q/R=~$constant or $\omega=\,$constant sections) belonging to the boundary of the parameter space that are of relevance by themselves. 
Examples of interesting boundary regions not mentioned in the preceding analysis are the surfaces $a/R=1$, $a/R=0$, and $\omega=1$.
Some properties of the objects belonging to the lines $a/R=1$ and $a/R=0$ were investigated in Ref.~\cite{Masa:2018elb} and then we do not reproduce such an analysis here. The properties of the solutions at the boundary $\omega=1$ are discussed in the next subsection.

\subsection{Regions in the parameter space}
\label{sec:regions}

\subsubsection{Preliminary remarks}

The boundaries described in the last subsection are surface boundaries of three-dimensional domains in the parameter space that contains objects with similar physical properties. In the two-dimensional sections of constant $q/R$, or of constant $\omega$, the surface boundaries appear as lines and the three-dimensional domains appear as two-dimensional regions.
As shown in Figs.~\ref{f:stab-waq=0}--\ref{f:stab-qaw=-1}, such regions indicate the different types of objects modeled by the solutions studied in the present work. White and light gray regions contain physically interesting objects. The hachured/gridded (grid with light brown dotted lines) regions present no real solutions since some of the parameters are complex numbers. A brief description of each region is giving in the following.

\subsubsection{  Region $(i)$}

This region contains regular undercharged star configurations with total gravitational mass larger than the total electric charge, $m/R>q/R$. 
The solutions in this region present a radius $a/R$ that satisfies the constraint $a/R>r_{+}/R$, meaning that the matching surface is outside the gravitational radius of the configuration and then no horizon is formed. 
All solutions are regular undercharged stars with a de Sitter core, a thin shell of positive mass ($M/R>0$), reassembling gravastars \cite{Mazur2001,Mazur:2004fk,Visser2004,BNNCarter:2005pi,Horvat:2008ch,Ghosh:2015ohi}. The configurations belonging to this region include from uncharged stars, as in the case of Fig.~\ref{f:stab-waq=0}, up to highly charged stars with $q/R$ very close to $m/R$. A configuration of this kind, i.e., with $q/R\lesssim m/R$, is singled out from any point of the parameter space within region $(i)$ that is located very close to one of the curves $c_1$ or $c_3$, as seen in the cases of Figs.~\ref{f:stab-waq=02}--\ref{f:stab-waq=078}, and also in all figures drawn in the $(q/R,\,a/R)$--plane, cf. Figs.~\ref{f:stab-qaw=1}--\ref{f:stab-qaw=-1}. In turn, the configurations that approach the gravastar limit, i.e., with shell radius $a/R$ very close to the normalized gravitational radius $r_+/R=(m+\sqrt{m^2-q^2})/R$, are the ones represented by points close to the boundary $c_3$, the ones for large negative values of $\omega$ (typically $\omega< - 0.8$), the configurations with parameter $a/R$ close to unity, and also the configurations with $\omega$ close to unity.

Note that there are two type $(i)$ regions in the parameter space. The two regions are present just for electric charges $q/R$ and parameter $\omega$ in the intervals $0\leq q/R\lesssim 0.7680$ and  $0<\omega \leq 1$, respectively. One of such regions is located above the boundary $c_{\sst\mp}$, while the other one is located below such a boundary. For electric charges in the interval $0.7680 \lesssim q/R<1$, as well as for $\omega\leq 0$, only the upper type $(i)$ region appears.  
Notice also that for sufficiently large electric charges the region $(i)$ is not present.
It can be shown that this class of undercharged (or uncharged, for $q/R=0$) stars satisfies the following constraint for the total electric charge, $0\leq {q}/{R}<1$, what may be verified by checking all the figures for $q/R\geq 1$. In fact, as seen from Figs.~\ref{f:stab-waq=1}--\ref{f:stab-waq=10}, region $(i)$ disappears for $q/R \geq 1$.

The boundary of the region $(i)$ located above the boundary line (surface in the 3D parameter space) $c_{\sst\mp}$ is formed by the lines (surfaces) $c_1$, $c_3$, $q/R=0$, $a/R=1$, and $\omega=1$, see Figs.~\ref{f:stab-waq=0}--\ref{f:stab-waq=078} and \ref{f:stab-qaw=1}--\ref{f:stab-qaw=-1}.
The boundary of the region $(i)$ located below $c_{\sst\mp}$ is formed by $c_1$, $c_3$, $q/R=0$, and $\omega=1$, see Figs.~\ref{f:stab-waq=0}--\ref{f:stab-waq=05775} and \ref{f:stab-qaw=1}--\ref{f:stab-qaw=-1}.
The case $q/R=0$ is special because in the limit $q/R\to 0$ the boundary $c_1$ is not present, it tends to the line $c_5$, which takes part in the region $(i)$ boundary, see Fig.~\ref{f:stab-waq=0}.

\subsubsection{ Region $(ii)$}

This region contains overcharged star configurations (no horizons are present) with total mass smaller than total electric charge,  $m/R< q/R$, except on the portion of line (surface) $c_3$ that crosses region $(ii)$ for which $q/R=m/R$, cf. Figs.~\ref{f:stab-waq=02}--\ref{f:stab-waq=078} and \ref{f:stab-qaw=1}--\ref{f:stab-qaw=-1}. 
As discussed above, the segment of $c_3$ inside the region $(ii)$ contains extremely charged ($q/R=m/R$) quasiblack holes.
All other solutions are regular overcharged stars with a de Sitter core, a thin shell of non-negative mass ($M/R\geq 0$), and with positive total mass ($m/R>0$). 

In the $(\omega,\, a/R)$--plane, different combinations of the lines $c_1$,  $c_2$, $c_4$, $c_{\sst{\mp}}$, $c_5$,  and $\omega=1$ that depend on the value of the electric charge form the boundary of this region, see Figs.~\ref{f:stab-waq=02}--\ref{f:stab-waq=10}. 

It is worth mentioning that $c_3$ does not belong to the boundary of the region $(ii)$, since in every situation the region continues across such line. In turn, $c_{\sst\mp}$ is a boundary of region $(ii)$ in some instances, e.g., between region $(ii)$ and region $(iv)$. In view of this difference, we consider that $c_{\sst\mp}$ is a boundary between two regions of type $(ii)$, cf. Figs.~\ref{f:stab-waq=02}--\ref{f:stab-qaw=-1}.

Naturally, region $(ii)$ is not present in the uncharged case of Fig.~\ref{f:stab-waq=0}. For values of electric charge and $\omega $ in the intervals $0< q/R < 3\sqrt{3}\,/4$ and $-1/2 <\omega\leq 1/2$, respectively,  the parameter space shows two type $(ii)$ regions, one of them above $c_{\sst\mp}$ and the other one below such a surface,  as seen in Figs.~\ref{f:stab-waq=02}--\ref{f:stab-waq=1}, and \ref{f:stab-qaw=1}--\ref{f:stab-qaw=-040}. The two regions $(ii)$ have a branch of $c_{\sst\mp}$ as the boundary between them.   As the electric charge increases from $q/R=1$, the upper region $(ii)$ in the $(\omega,\, a/R)$--plane shrinks down to vanish at $q/R=3\sqrt{3}\,/4$. As seen from Figs.~\ref{f:stab-waq=1.3}--\ref{f:stab-waq=10}, the lower region $(ii)$ also tends to disappear for large $q/R$.

In the ($q/R$, $a/R$)--plane, the boundary of region $(ii)$ is formed by different combinations of the lines $c_1$,  $c_2$, $c_4$, $c_{\sst{\mp}}$, $c_5$,  and $a/R=0$, as seen in Figs.~\ref{f:stab-qaw=1}--\ref{f:stab-qaw=-1}.
As also seen in those figures, the line $c_{\sst\mp}$ is not present for all $\omega \leq -1/2$ and we are left with only one region of type $(ii)$.

\subsubsection{Region $(iii)$}

This region is the most relevant for our purposes. All objects contained in such a region satisfy the constraint $a/R<r_{-}/R$, where $r_-$ is the Cauchy horizon, confirming they are all charged regular black holes. The central core is a regular distribution of charged fluid whose radial pressure satisfies a de Sitter equation of state, and whose boundary is a thin shell located at $a/R< r_-/R$. The spacetime metric in the region $r>a$ is the RN electrovacuum solution.
In cases where the boundary $c_2$ belongs to the frontier of region $(iii)$, we find regular black hole configurations with a massless shell exactly on that line,
i.e., the mass of the shell at the boundary of the object vanishes and the matching between the inner and the outer metrics is smooth, by means of a boundary surface. This happens for electric charges in the interval $3\sqrt{3}\,/4 \leq q/R< \sqrt{3}$ for all $\omega$, as shown in the cases of Figs.~\ref{f:stab-waq=1.3} and \ref{f:stab-waq=1.5}. The presence of $c_2$ at the boundary of the type $(iii)$ region located above the $c_{\sst\mp}$ is clearly seen in Figs.~\ref{f:stab-waq=1.3}--\ref{f:stab-waq=1.5} and \ref{f:stab-qaw=1}--\ref{f:stab-qaw=-1}.

Region $(iii)$ does not appear in the $q/R=0$ case, see Fig.~\ref{f:stab-waq=0}.
Two regions of this kind are present in all the domain with electric charge $q/R$ and $\omega$ in the intervals $0<q/R< \sqrt{3}$ and $ 0< \omega \leq 1$, respectively. One of such regions is above (and to the left of) and another one is below (and to the right of) the boundary $c_{\sst\mp}$, as seen in Figs.~\ref{f:stab-waq=02}--\ref{f:stab-waq=1.5} and \ref{f:stab-qaw=1}--\ref{f:stab-qaw=015}. In the $(\omega,\, a/R)$--plane, for $q/R\geq \sqrt{3}$ only the region $(iii)$ located below $c_{\sst\mp}$ is present, cf. Figs.~\ref{f:stab-waq=sqrt3}--\ref{f:stab-waq=10}. On the other hand, in the $(q/R,\, a/R)$--plane, for $\omega\leq 0$ only the region located above the $c_{\sst\mp}$ is present, cf. Figs.~\ref{f:stab-qaw=0}--\ref{f:stab-qaw=-1}.

The boundary of the two regions of type $(iii)$ varies along the three dimensional parameter space. One of such regions is bounded by branches of the surfaces $a/R=0$, $c_1$ (or $c_2$), $c_3$ (or $a/R=1$), and $\omega=1$. The other one is bounded by branches of the surfaces $c_1$, $c_2$, $c_3$ (or $c_{\sst\mp}$), and $\omega=1$. The physical properties of the objects belonging to each different branch of this frontier were presented in Sec.~\ref{sec:boundaries}.

\subsubsection{Region $(iv)$}

This is another region that contains regular overcharged stars (no horizon is present) with total positive mass ($0< m/R<q/R$). The main difference when compared to region $(ii)$ is that the mass of the thin shell is negative ($M<0$). A single point from this region with specific values of $m/R$, $q/R$ and $a/R$ represents a spacetime whose geometric properties are basically the same as a configuration singled out from region $(ii)$.  There is a central core of charged fluid whose radial pressure satisfies a de Sitter equation of state, and whose boundary is a thin shell located at $a>r_+$, where $r_+$ is the gravitational radius of the solution.

Region $(iv)$ is not present in the boundaries of zero charge $q/R=0$ and for $\omega=1$, and also for the especial case with $q/R=1/\sqrt{3}$.
There are two of such regions for $\omega$ in the interval $-1/2 <\omega <0$, while there is just one of such regions for $\omega$ in the intervals $1>\omega\geq0$ and $-3/2<\omega\leq -1/2$. Moreover, this type of region is not present in the case $\omega=1$ and in the interval $\omega<-3/2$ (see Figs.~\ref{f:stab-qaw=1}--\ref{f:stab-qaw=-1}).

As seen in Figs.~\ref{f:stab-waq=1.5} and \ref{f:stab-waq=10}, region $(iv)$ tends to disappear, becoming vanishingly tiny, for large values of $q/R$. For sufficiently small electric charge, $0<q/R \leq 3\sqrt{3}\,/4$, the region is delimited by $c_2$, $c_{\sst\mp}$, and a branch of $c_5$, while for large electric charges, with $q/R>3\sqrt{3}\,/4$, it is bounded by $c_1$, $c_{\sst\mp}$, and $c_5$.
In the $(q/R,\, a/R)$--plane, depending on the values of $\omega$, region $(iv)$ is delimited by the lines $c_1$, $c_2$, $c_{\sst\mp}$, $c_4$, and a branch of $c_5$.

\subsubsection{Region $(v)$}

This is another region that contains regular charged black hole solutions with two horizons, i.e., with total gravitational mass larger than the electric charge  $m/R> q/R$, whose central core of charged matter is bounded by a thin shell located inside the Cauchy horizon, at the radius $a/R< r_-/R$. The main difference with respect to the objects found in region $(iii)$ is that the mass of the shell $M/R$ is negative. 
A single point from this region with specific values of $m/R$, $q/R$ and $a/R$ represents a spacetime whose geometric properties are basically the same as a configuration singled out from region $(iii)$. 

In the ($\omega$, $a/R$)--plane, the region shows up just for large values of electric charge, $q/R>3\sqrt{3}\,/4$. For electric charges in the interval $3 \sqrt{3}/4<q/R<\sqrt{3}$, it is delimited by the lines $c_1$, $c_2$, $c_{\sst\mp}$, and $\omega=1$ (see Fig.~\ref{f:stab-waq=1.5}), while for $q/R\geq \sqrt{3}$, it is delimited by the lines $c_1$, $c_{\sst\mp}$, $a/R=1$, and $\omega=1$ (see Figs.~\ref{f:stab-waq=sqrt3}--\ref{f:stab-waq=10}).
In the ($q/R$, $a/R$)--plane, the region is present for all $\omega$, and it is delimited by the lines $c_1$, $c_2$, $a/R=1$, and $c_{\sst\mp}$,  see Figs.~\ref{f:stab-qaw=1}--\ref{f:stab-qaw=-1}.

\subsubsection{Region $(vi)$}

This region contains regular objects without horizons resembling regular stars, but with negative total gravitational mass, $m/R<0$, and with a thin shell of positive mass, $M/R>0$. These kinds of configurations appear for $\omega$ in the interval $-1/2<\omega<0$, and two regions of this type show up in some cases. The upper region, present just for small charges ($0\leq q/R<1/\sqrt{3}$), is delimited by the lines $c_{\sst\mp}$, $c_4$, and a branch of $c_5$, see Figs.~\ref{f:stab-waq=0}--\ref{f:stab-waq=02} and \ref{f:stab-qaw=-022}--\ref{f:stab-qaw=-040}. The lower region is bounded by  $a/R=0$, $c_{\sst\mp}$, $c_4$, and a branch of $c_5$.  In view of the total gravitational mass being negative, the solutions in this region are of little interest.

\subsubsection{Region $(vii)$}

This region contains regular objects with no horizon resembling regular stars, but with negative total gravitational mass $m/R<0$, and with a thin shell also with negative mass, $M/R<0$, and then the solutions in this region are of little interest. The region is delimited by the lines $c_4$, $c_5$, and $c_{\sst\mp}$, or $q/R=0$, $c_4$, $c_{\sst\mp}$, and $a/R=0$ at the frontier $q/R=0$, see Figs.~\ref{f:stab-waq=0}--\ref{f:stab-qaw=-1}.

\subsubsection{Region $(viii)$}
This is the region with no real solution for $M/R$, i.e., $M/R$ assumes complex values meaning that there in no interest in these configurations. The region is delimited by the lines $a/R=0$ and $c_{4}$, and/or by the lines $c_4$ and $\omega=1$, see Figs.~\ref{f:stab-waq=0}--\ref{f:stab-qaw=-1}.

\section{Stability analysis of the equilibrium solutions}\label{sec:4}

\subsection{General remarks}To investigate the stability of the thin shell against radial perturbations about the static solution $a=a_0=~$constant, it is useful rewriting the equation for the surface energy density $\sigma$, Eq.~\eqref{eq:eds}, in the following suggestive form~\cite{Lake1979} 
\begin{equation}\label{eq:motionequation}
\dot{a}^{2}+V(a)=0,
\end{equation}
where now $a$ is a time dependent variable $a=a(\tau)$.
Taking cognizance of Eqs.~\eqref{eq:eos} and \eqref{eq:EdTS}, the effective potential $V(a)$ may be written as 
\begin{multline}\label{eq:PotenEq2}
V(a)=-\left[ 
\frac{1}{2M_{0}}\left(\frac{a^{3}}{R^{2}}+\frac{q^{2}}{a}-2m\right)\left(\frac
{a_{0}}{a}\right)^{-2\omega}\right. \\
-\left. \frac{M_{0}}{2a_{0}}\left(\frac{a_{0}}{a}\right)^{1+2\omega}\right] ^{2}-\dfrac{a^{2}}{R^{2}}+1,
\end{multline}
where $M_{0}=4\pi a_{0}^{2}\sigma(a_{0})$ is a constant representing the mass of the shell at equilibrium.

Here, the equilibrium (static) solution $a =a_{0}$ can be obtained by solving simultaneously the equations $V(a_{0})=0$ and $V^{\prime}(a_{0})=0$, which means that the configuration is at rest, i.e., $\dot{a}=\ddot{a}=0$. The relations $V(a_{0})=0$ and $V^{\prime}(a_{0})=0$ lead respectively to expressions~\eqref{eq:massadashell4} and \eqref{eq:totalmass}.

Now, in order to obtain the stability conditions of the static solutions, we consider a Taylor expansion of the effective potential $V(a)$ around $a_ {0}$, 
\begin{multline} \label{eq:Poten3}
V(a)=V(a_{0})+V^{\prime}(a_{0})(a-a_{0})\\
+\frac{1}{2}V^{\prime\prime}(a_{0})(a-a_{0})^{2}+\mathcal{O}\left[\left(a-a_{0}\right)^{3}\right].
\end{multline}
By substituting the equilibrium conditions into Eq.~\eqref{eq:Poten3}, it follows
\begin{equation}
V(a)=\frac{1}{2}V^{\prime\prime}(a_{0})(a-a_{0})^{2}+\mathcal{O}\left[
\left(a-a_{0}\right)^{3}\right].
\end{equation}
Hence, the stability condition may be stated as follows.  If $V(a)$ has a local minimum at $a=a_{0}$ and $V^{\prime\prime}(a_{0})>0$, the solution at $a=a_0$ is stable. On the other hand, the condition $V^{\prime\prime}(a_{0})<0$ implies instability of the thin shell. If $V^{\prime\prime}(a_{0})=0$, the present criterion is inconclusive, and then the next nonzero $n$-derivative of $V(a)$ is necessary to characterize the potential and to define unambiguous stability conditions. This particular situation is not considered here.

The second derivative of the potential evaluated at the static point $a_{0}$ is obtained from relation \eqref{eq:PotenEq2}, which gives
\begin{widetext}
\begin{multline}\label{eq:derivada2V}
\dfrac{1}{2}V^{\prime\prime}(a_{0}) =-\dfrac{1}{R^{2}} -\left[\dfrac{\left( 3+2\omega\right)a_{0}^{2}}{2M_{0}R^2}- \frac{(1-2\omega)q^{2}}{2M_0\, a_0^2} -\dfrac{2m\,\omega}{M_{0}\,a_{0}}+\dfrac{(1+2\omega)M_{0}}{2a_{0}^{2}} \right]^{2} -\left[\dfrac{ a_{0}^{4}+\left(q^{2}-2m\,a_{0}\right)R^2}{2M_{0}\,a_{0}R^{2}} \right.\\  
\left. - \dfrac{M_{0}}{2a_{0}} \right] \left[\dfrac{\left(1+\omega\right)\left(3+2\omega\right)a_0}{M_{0}R^{2}}
+\frac{\left(1+\omega\right)\left(1-2\omega\right)q^{2}}{M_{0}\,a_{0}^{3}}
+\frac{2\left(1-2\omega\right)m\,\omega}{M_0\, a_{0}^{2}} +\dfrac{\left(1+\omega\right)\left(1+2\omega\right)M_{0}}{a_{0}^{3}}\right],
\end{multline}
\end{widetext}
where  $M_{0}$ and $m$ are obtained putting $a=a_0$ in  Eqs.~\eqref{eq:massadashell4} and \eqref{eq:totalmass}, respectively.

The stability analysis of spherical thin shells, isolated or in the presence of a central compact object, by following the strategy just reviewed has been widely employed in the literature, see. e.g. Refs. \cite{Ishak:2001az,Lobo:2005zu,Dias:2010uh,Eiroa:2011nd,Varela2015} and their references for a small sample of such kind of works, see also Refs.~\cite{Rosa:2020hex,Alestas:2020wwa} for more recent works, and the very interesting work of Ref.~\cite{PerezBergliaffa:2020zzv} for other kind of analysis and more references on the subject.

The ingredients for the stability analysis are now ready.
Since the model presents three free parameters, namely $a_0/R$, $q/R$, and $\omega$, the condition $V^{\prime\prime}(a_{0})=0$ defines a surface in the corresponding parameter space. Such a surface separates the space into disjoint regions containing only stable, or only unstable solutions. For a better visualization of such regions we perform the analysis by slicing the parameter space first considering the planes with constant $q/R$, and then the planes of constant $\omega$.

\subsection{Regions of stability in the \boldmath{$(\omega,\, a/R)$}--plane}

\subsubsection{General remarks}

Here we investigate the stability of the solutions by slicing the parameter space at some fixed values of $q/R$ and determining the values of $a_{0}/R$ and $\omega$ for which $V^{\prime\prime}(a_{0})=0$. Such an equation defines a curve (or a set of curves) in the two dimensional slice of the parameter space that separates the planar slice into regions containing stable configurations from regions containing unstable configurations. The study is performed for a few different values of $q/R$, and the results are presented in the set of graphs shown in Figs.~\ref{f:stab-waq=0}--\ref{f:stab-waq=10}. A brief description of the physical properties of the corresponding stable (unstable) solutions is given below. In the remaining of this section and in the labels of all figures, to simplify notation, we drop the index ``0'' by identifying $a_0\equiv a$, $M_0\equiv M$, etc.

\subsubsection{The zero electric charge case, $q/R=0$} 

\begin{figure}[h]
\centering 
\includegraphics[width=0.37\textwidth]{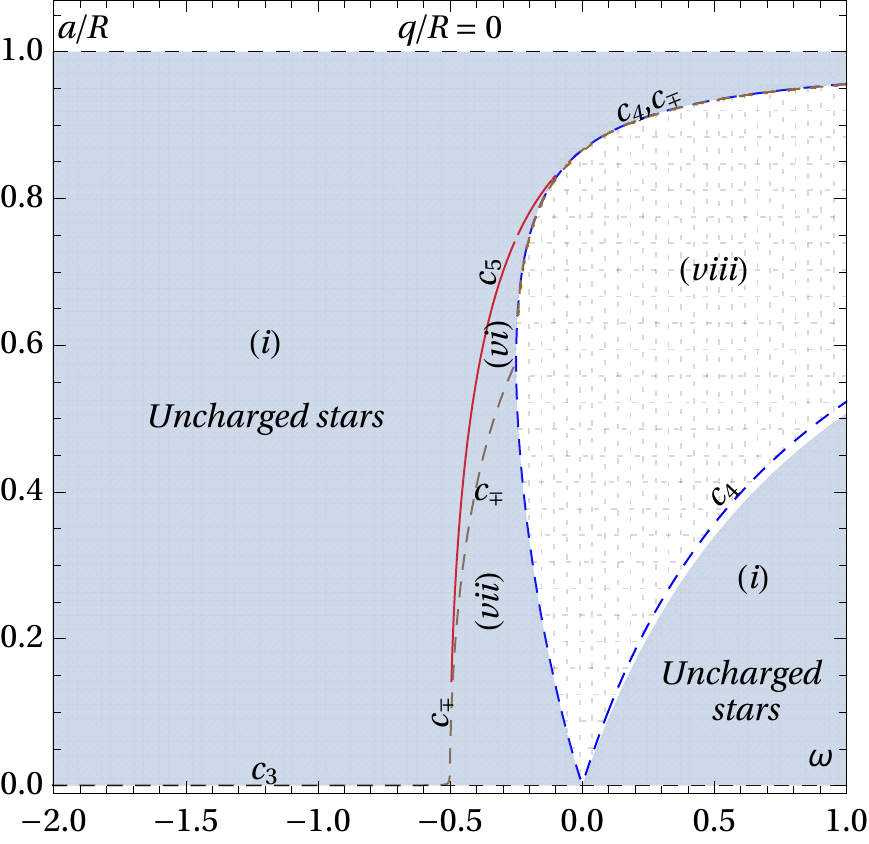}
\caption{The stability (white) and instability (light gray) regions for $q/R=0$ in the $(\omega,\, a/R)$--plane. The gridded region $ (viii)$ contains no physical solutions.}
\label{f:stab-waq=0} 
\end{figure}

Figure~\ref{f:stab-waq=0} shows the regions containing stable and unstable uncharged ($q/R=0)$ solutions in the parameter space. The figure is drawn in the $(\omega,\, a/R)$--plane, and $q/R =0$ is a boundary surface of the three-dimensional parameter space. The white regions represent stable solutions and the light gray regions correspond to the unstable solutions, while the gridded region $(viii)$ contains no physical solutions and is not considered in the present analysis.

In this uncharged case, we find stable objects just in region $(i)$ which, as described in Sec.~\ref{sec:regions}, are uncharged regular stars with positive total gravitational mass $m/R> 0$. These stable stars are found for $\omega$ in the interval  $0<\omega\leq 1$, in a slim region just below the curve $c_4$, with the parameter $a$ varying within the interval $0<a/R\lesssim 0.5230$.

It is worth mentioning that the particular case with $q/R=0$ and $\omega=1$ corresponds to a particular interesting case of the model for stable gravastars studied in Ref.~\cite{Visser2004}, see also \cite{Chirenti:2007mk} for a different stability analysis of gravastar models. 
In the critical case of~\cite{Visser2004}, the condition for the existence of (stable) thin shells which satisfy the stiff equation of state (${\cal P} =\sigma$) implies the constraint $k\,m^{2}\simeq 0.02430$, with $k$ being a constant, see Eq.~(63) in Ref.~\cite{Visser2004}. In our notation, $k$ is given by $k=1/2R^{2}$. It is then found that stable configurations occur if the total mass is smaller than or equal to a critical value given by $m_c/R\simeq 0.2205$ (our notation). For such a critical mass, the thin shell is located at $a/R\simeq 2.301 m_c/R \simeq 0.5072$, implying in $a/r_h=a/2m=1.150$.
As it can be verified in the case of Fig.~\ref{f:stab-waq=0}, in our model the thin shell stability occurs for $a/R$ in the interval  $0.5072\lesssim a/R\lesssim 0.5230$, corresponding respectively to total gravitational masses in the range $0.2205\gtrsim m/R\gtrsim 0.2092$. The ratio between the thin shell radius and the gravitational radius ($2m$) is in the interval $1.150 \lesssim a/2m \lesssim 1.260$, where the lower limit is in agreement with the results of Ref.~\cite{Visser2004}. 

 A true gravastar configuration presents a boundary which is very close to the corresponding gravitational radius, so that even small arbitrary perturbations on the position of the boundary may lead to the shell to reach the gravitational radius, giving rise to an event horizon and forming a black hole \cite{Reviewer}.
Roughly speaking, an equilibrium solution may suffer a perturbation $\delta a$ as large as the {\it distance} between the initial radius of the shell $a=a_0$ and the corresponding gravitational radius of the configuration $r_+= m +\sqrt{m^2-q^2} =2m$ (in the case $q/R=0$), without reaching the corresponding gravitational radius.
As we have just mentioned, taking $\omega=1$ as a representative case, stable gravastars occur for  $0.5072\lesssim a/R\lesssim 0.5230$, with the ratio $a/2m$ respectively in the range $1.150 \lesssim a/2m \lesssim 1.260$. Therefore, keeping fixed all the other free parameters, the relative amplitude of perturbations on the shell position ($\delta a/2m$) may be as large as $0.15$, which means 15\% of its equilibrium relative value ($a_0/2m_0$).  
In fact, the junction conditions are satisfied all along during the shell oscillations, and it can be shown that the gravastar configurations enter an instability region of the parameter space before the ratio $a/2m$ reaches the limiting value $a/2m=1$.
With such initial conditions, in principle, the system will oscillate around the equilibrium configuration.

\subsubsection{The case with $q/R=0.20$} 

\begin{figure}[h]
\centering 
\includegraphics[width=0.37\textwidth]{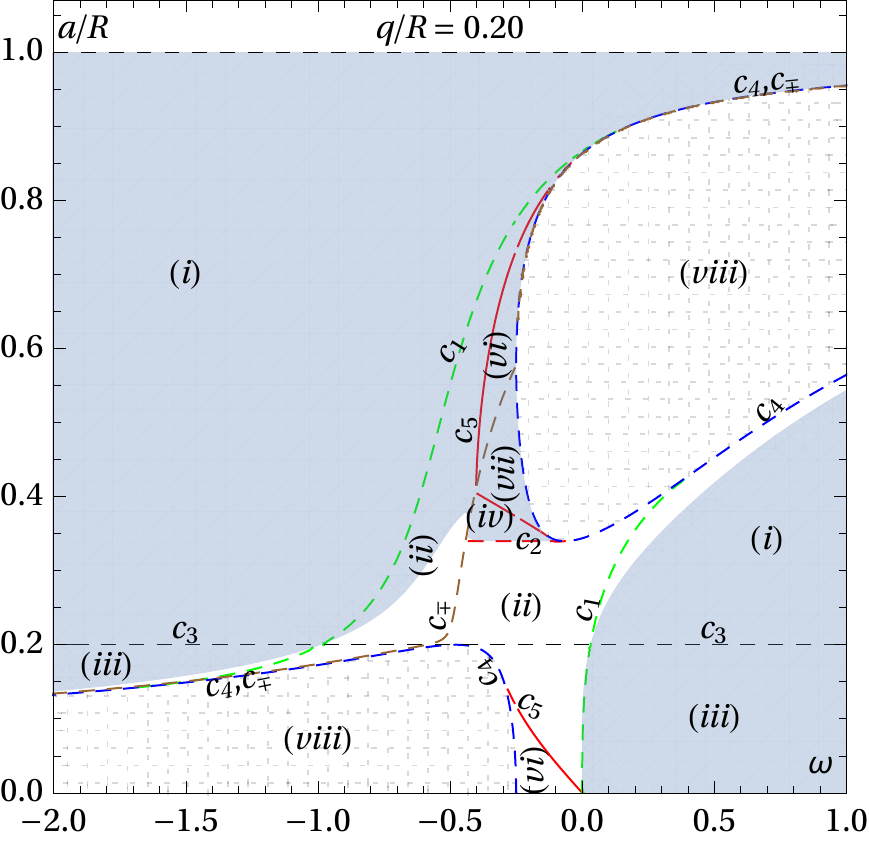}
\caption{Stability (white) and instability (light gray) regions for $q/R=0.2$ in the $(\omega,\, a/R)$--plane. 
The gridded regions $(viii)$ contain no physical solutions. }
\label{f:stab-waq=02} 
\end{figure}

This case, whose results of the stability analysis are shown in Fig.~\ref{f:stab-waq=02}, is representative of all configurations with electric charge in the interval $0< q/R < 1/\sqrt{3}$. As in the other figures presented in this section, the white regions represent stable solutions and the light gray regions contain unstable solutions in the $(\omega,\, a/R)$--plane, while the gridded regions  $(viii)$ contain no physical solutions.
According to the figure, there are stable objects of four different types, in the regions $(i)$, $(ii)$, $(iii)$, and $(vi)$.

A small part of the undercharged stars belonging to the region $(i)$ with parameters $a$ and $\omega$ respectively in the intervals $0.20\leq a/R \lesssim 0.5643$ and
$0.03114\lesssim\omega\leq 1$ are stable.  The range of the total gravitational masses of these stable charged stars is $0.20\leq m/R \lesssim 0.2823$, with the compactness ratio $m/a$ in the interval $0.4754 \lesssim m/a\leq 1 $, where the least compact objects correspond to $a/R\simeq 0.5643$ and $m/R\simeq 0.2683$, beyond the Buchdahl limit established for uncharged spheres~\cite{Buchdahl} and in accordance with the analog limit for charged spheres~\cite{Andreasson09,Lemos:2015wfa}.  As mentioned above, configurations in region $(i)$ may be interpreted as charged gravastars, and then these configurations represent stable gravastars, which are good black hole mimickers. As in the preceding case, for $q/R=0$, the ratio between the thin shell radius $a$ and the gravitational radius ($r_{+}=m+\sqrt{m^2-q^2}$) depends on the state equation of matter on the shell, with the most interesting stable gravastar configurations occurring for stiff matter, with $\omega =1$. 
In such a case, stable gravastars are found for $a/R$ in the interval $0.5441\lesssim a/R \lesssim 0.5643$, corresponding respectively to total gravitational masses in the range $0.2822\gtrsim m/R\gtrsim 0.2683$, and  with the ratio $a/r_{+}$ is in the interval $1.130\lesssim a/r_{+} \lesssim 1.256$. Therefore, by keeping fixed all the other free parameters, the amplitude of perturbations of the (stiff matter) shell position may be as large as 13\% of its equilibrium relative value $a_0/r_+$ without reaching the corresponding gravitational radius. In fact, as in the uncharged case of Fig.~\ref{f:stab-waq=0}, it can be shown that the oscillating gravastar configurations enter a neighboring unstable portion of region $(i)$  before the ratio $a/r_+$ reaches the limiting value $a/r_+=1$. In case of charged gravastars, the oscillating configuration may otherwise enter a type $(ii)$  region, by crossing the line $c_1$, and turning into a stable overcharged star.

All overcharged stars present in the regions $(ii)$ on the right of line $c_{\sst\mp}$, for which $-0.4347\lesssim \omega\lesssim 0.4269$ and $0<a/R\lesssim 0.4342$, and part of configurations in the region on the left of such a curve, for which $-1.709\lesssim\omega \lesssim -0.4347$ with $0.1415 \lesssim a/R\lesssim 0.3853$,  are stable solutions against radial perturbations of the shell. The range of the total gravitational masses of these stable configurations is $0\leq m/R < 0.2$, with the maximum mass configurations being located very close the line $c_1$, and the minimum mass at line $c_5$. In fact, the overcharged stars with zero total mass $(m/R=0)$ represented by points on the branch of line $c_5$ located below line $c_3$ are also stable configurations.

Another interesting kind of stable ultracompact objects are the extreme quasiblack holes found on the segment of the line $c_3$ that is inside the region $(ii)$. That is in the interval $-0.9895\lesssim\omega\lesssim 0.03114$, the masses of these objects equal the corresponding electric charges $m/R=q/R=0.2$, and the shell is located at $a/R=q/R=0.2$. From the point of view of an external observer, the matching surface is at the extreme RN horizon, $a/R=r_-/R=r_+/R$, so that each configuration on the mentioned segment of $c_3$ corresponds to a stable quasiblack hole.

More interestingly, in this case with $q/R=0.2$ it happens part of the regular black hole configurations belonging to one of regions $(iii)$ are stable solutions. These stable regular black holes are found in the region given by $\omega$ and $a/R$ in the intervals   
 $\omega\lesssim -0.9895 $ and $0< a/R < 0.2$, respectively, a region that becomes vanishingly thin as $\omega$ decreases to large negative values. The range of gravitational masses of these configurations is very close to the extreme solution $m/R\simeq q/R= 0.2$. 
Note that the extreme regular black holes found on the segment of the line $c_1$ located between the curves $c_3$ and $c_4$, with $\omega$ in the interval $-1.709\lesssim\omega\lesssim-0.9895$, are also stable configurations. 

Other stable objects are the regular charged stars found in the branch of region $(vi)$ located below the line $c_3$. This kind of solutions represent less interesting objects than the other regions due to the fact that they carry negative total gravitational mass.

\subsubsection{The case with $q/R=1/\sqrt{3}$}

\begin{figure}[h]
\centering 
\includegraphics[width=0.37\textwidth]{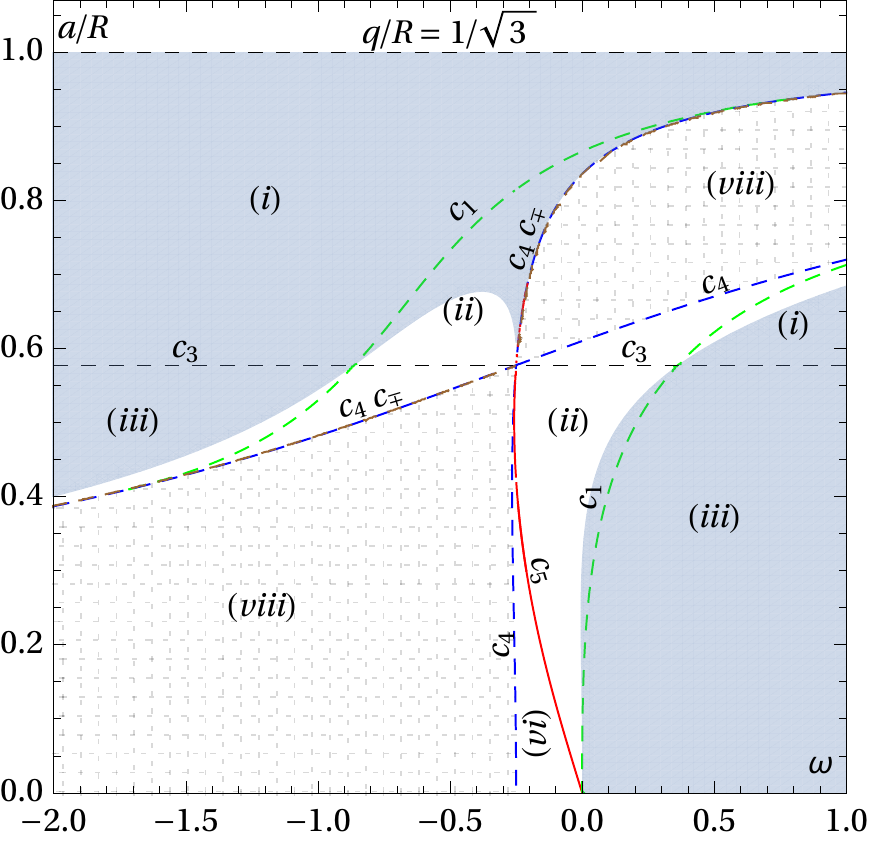}
\caption{Stability (white) and instability (light gray) regions for $q/R= 1/\sqrt{3}\simeq 0.57735$ in the $(\omega,\, a/R)$--plane. 
The gridded regions $(viii)$ contain no physical solutions. }
\label{f:stab-waq=05775} 
\end{figure}

Figure~\ref{f:stab-waq=05775} shows the results of the stability analysis for $q/R=1/\sqrt{3}$ in the $(\omega,\, a/R)$--plane. 
This case is chosen because it shows a particular feature. The two branches of the curve $c_4$ share the point with coordinates ($\omega=-1/4$, $a/R=1/\sqrt{3}$), and the whole line $c_{\sst\mp}$ coincides with the upper branch of $c_4$. As a consequence, the line $c_2$, the upper branch of line $c_5$, the region $(iv)$, the upper region $(vi)$, and the region $(vii)$ are not present.
As in the case of Fig~\ref{f:stab-waq=02}, stable solutions are found in four different regions, namely, in the white portions of regions $(i)$, $(ii)$, $(iii)$, and $(vi)$.

A small portion of the region $(i)$ located below $c_4$, between the curves $c_1$ and $c_3$, with $\omega$ in the interval $0.3624\lesssim\omega\leq 1$, contains stable regular undercharged stars, also interpreted as charged gravastars. The range of gravitational masses of these stable stars is $1/\sqrt{3} < m/R\lesssim 0.5812$, with the radius in the interval  $1/\sqrt{3} < a/R\lesssim 0.7128$, and the compactness factor in the interval $0.8100\lesssim m/a < 1$.
 In the case of the gravastars with stiff matter on the shell, stable configurations occur for total mass in the interval $0.5812\gtrsim m/R\gtrsim0.5773$, for $a/R$ respectively in the interval $0.6851\lesssim a/R \lesssim 0.7199$, and with the compactness ratio $a/r_{+}$ in the range $1.056\lesssim a/r_{+} \lesssim 1.231$. 
 In this case the amplitude of perturbations of the (stiff matter) shell position may be as large as 5\% of its equilibrium relative value $a_0/r_+$ without reaching the corresponding gravitational radius. As in the previous cases, it can be shown that, before the ratio $a/r_+$ reaches the limiting value $a/r_+=1$, oscillating charged gravastar configurations enter either a neighboring unstable type $(i)$ region or a stable type $(ii)$ region, changing into an overcharged star.

Almost all the configurations belonging to the branches of region $(ii)$ which lie below the curve $c_4$, in the region with $-1/4\leq \omega \leq 1$, are stable overcharged stars. 
A great portion of the other branch of region $(ii)$ which lies above the curve $c_4$ with $\omega$ in the interval $-1.720\lesssim\omega<-1/4$ also contains stable overcharged stars. 

Configurations represented by the line $c_5$ are stable electrically charged object with zero gravitational mass. 

Stable extreme quasiblack holes configurations are found on line $c_3$ for $-0.8624\lesssim\omega\lesssim0.3624$. The mass of each one of such objects equals the electric charge $m/R=q/R=1/\sqrt{3}$, and the shell is located at $a/R=r_-/R=r_+/R=1/\sqrt{3}$.

A portion of region $(iii)$, for $\omega$ and $a/R$ in the ranges $\omega\lesssim -0.8624$ and $0< a/R< 1/\sqrt{3}$, contains stable regular black holes. The stable region vanishes as $\omega$ decreases to high negative values.
Note also that the extreme regular black holes found on the line $c_1$, for $\omega\lesssim -0.8624$, are stable solutions too.

The entire region $(vi)$ presents stable configurations with negative gravitational mass.

\subsubsection{The case with $q/R=0.78$} 

\begin{figure}[h]
\centering 
\includegraphics[width=0.37\textwidth]{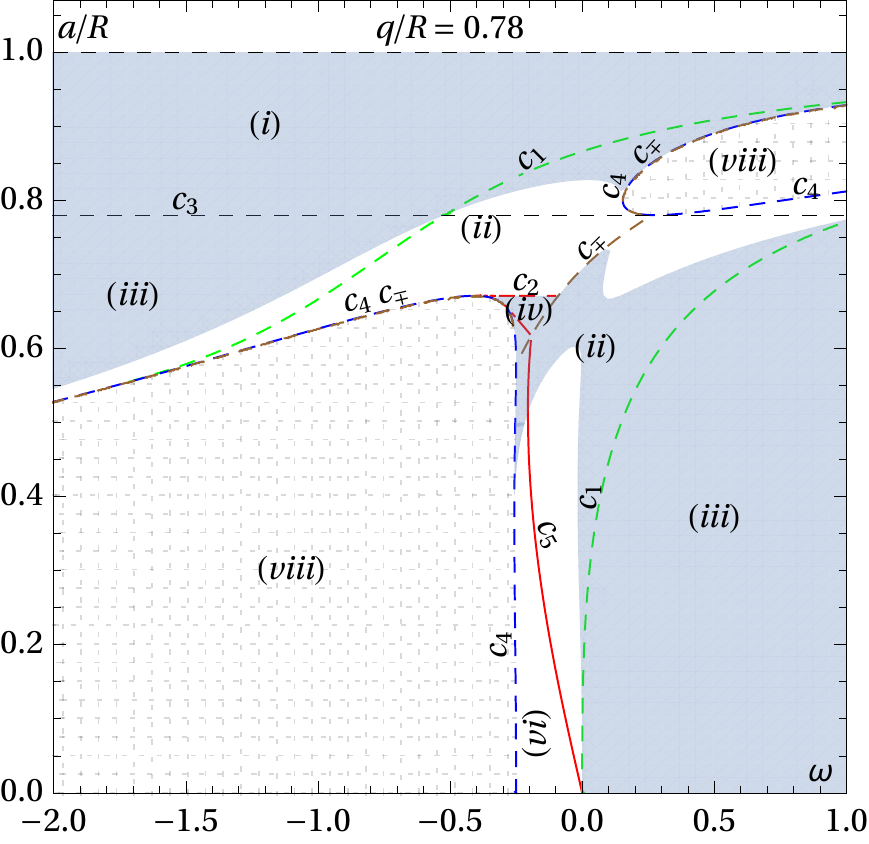}
\caption{Stability (white) and instability (light gray) regions for $q/R=0.78$ in the $(\omega,\, a/R)$--plane. Region $(vii)$ is also present, it is just below region $(iv)$, on top of region $(vi)$, but it is not indicated by a label in the figure.
The gridded regions $(viii)$ contain no physical solutions. }
\label{f:stab-waq=078} 
\end{figure}

This case is chosen as representative of all instances with electric charge in the interval $0.7680\lesssim q/R< 1$, for which there is only one region of type $(i)$, i.e., the region $(i)$ that for smaller values of charge appears below the curve $c_4$, between the lower branch of $c_1$ and the line $c_3$, is not present here. In this case, as it happens in all cases with electric charge  in the mentioned interval, the curve $c_2$, the regions $(iv)$, $(vi)$ and $(vii)$ appear only below the curve $c_3$.
The results of the stability analysis for the case $q/R=0.78$ in the $(\omega,\, a/R)$--plane are presented in Fig.~\ref{f:stab-waq=078}. As in the previous cases, the white regions represent stable solutions and the light gray regions are the unstable solutions. Here, we can see stable solutions in the regions $(ii)$, $(iii)$ and $(vi)$.

Differently form the preceding cases, there is no stable undercharged stars (gravastars) in type $(i)$ regions, since the region $(i)$ located above $c_4$ does not present stable configurations.  

Region $(ii)$ shows stable regular overcharged stars in a large sub-region in the parameter space, i.e., the stable region is bounded from below by branches of the curves $c_2$ and $c_4$, and from above, in part, by a branch of curve $c_1$ and other branch of $c_4$. The extreme values of the parameters are   $-1.7249\lesssim \omega\leq 1$ and $0 < a/R\lesssim 0.8273$.

Stable quasiblack holes solutions are found on the segment of line $c_3$ that crosses region $(ii)$, for $-0.5222\lesssim\omega\leq 1$. These objects have parameters satisfying the relations $m/R=q/R=r_-/R=r_+/R=a/R=0.78$.

Region $(iii)$ contains stable charged regular black holes in a slim area just above the curve $c_1$. The intervals of parameters are $\omega\lesssim -0.5222$ and $ a/R < 0.7800$, with the stable region becoming vanishingly thin as $\omega$ reaches high negative values. Since the stable region is close to $c_1$, the masses of such black holes are just slightly higher than the corresponding electric charges.
Stable extremely charged regular black holes are found on the segment of line $c_1$ that is between regions $(ii)$ and $(iii)$, for $\omega$ in the interval $-1.7249\lesssim\omega\lesssim -0.5222$.
 
 Overcharged stars with zero total mass $(m/R=0)$ on a segment of line $c_5$, for $\omega$ in the interval $-0.2056\lesssim\omega < 0$,  are stable.
 
 The stable portion of region $(vi)$ is for the range of parameters  $-0.25\lesssim\omega < 0$ and $0 < a/R\lesssim 0.5135$.

\subsubsection{The case with $q/R=1$}

\begin{figure}[h]
\centering 
\includegraphics[width=0.37\textwidth]{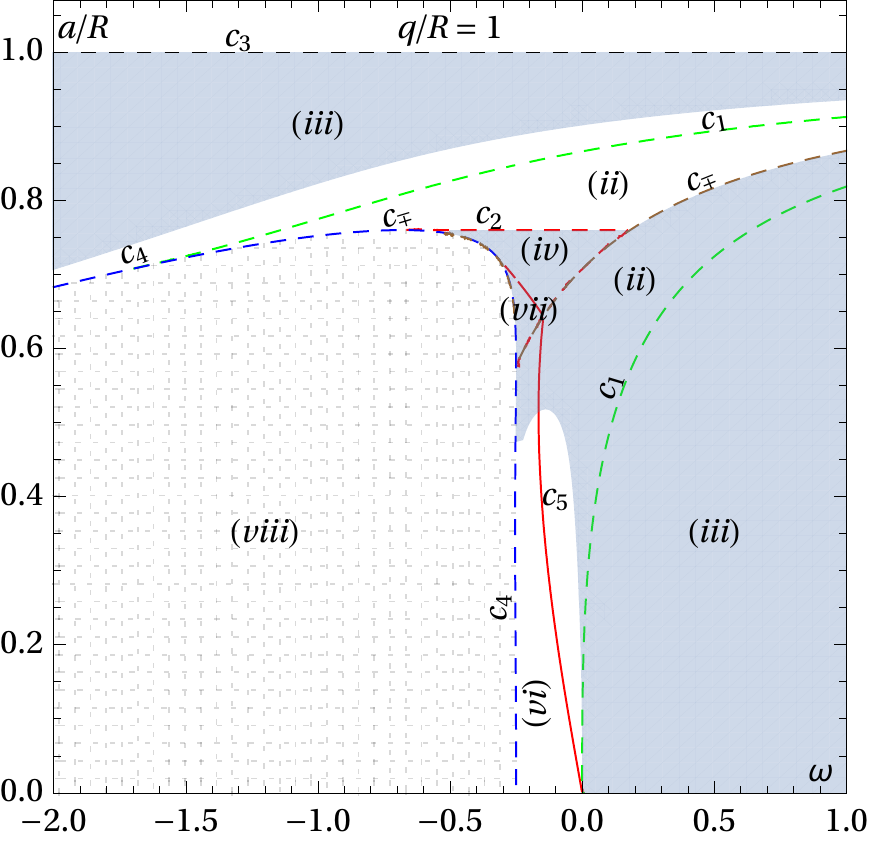}
\caption{Stability (white) and instability (light gray) regions for $q/R=1$ in the $(\omega,\, a/R)$--plane. 
The gridded region $(viii)$ contains no physical solutions. }
\label{f:stab-waq=1} 
\end{figure}

Figure~\ref{f:stab-waq=1} shows the results of the stability analysis for $q/R=1$ in the $(\omega,\, a/R)$--plane. 
This case is chosen because it shows a few particular different features in comparison to the cases $0<q/R<1$. The region $(i)$ of undercharged stars disappears for $q/R\geq 1$, the branch of $c_4$ lying in the positive region of $\omega$ is not present, and neither the corresponding region $(viii)$. The line $c_3$ also does not appear, except for the branch that coincides with the line $a/R=1$. 
As in the case with $q/R=0.78$, here the stable solutions are found in the regions  $(ii)$, $(iii)$, and $(vi)$.

The whole region $(ii)$ above $c_\mp$ and a portion of $(ii)$ below such a line, with parameters $\omega$ and $a/R$ in the intervals $-0.1653\lesssim\omega\leq 0$ and $0< a/R\lesssim 0.5177$, show stable regular overcharged stars. The configurations close to curve $c_2$ present stable thin shells with small intrinsic mass close to zero. The present stability criterion fails for configurations on the line $c_2$, since the second derivative of the potential, cf. Eq.~\eqref{eq:derivada2V}, is not defined there. Within the white portion of the lower region $(ii)$, the configurations located next to the line $c_5$ are stable overcharged stars with small mass compared to the electric charge ($m/R\sim 0)$, while the configurations located next to the line $c_1$ have mass close to the electric charge ($m/R\sim 1$).

The region of stable charged regular black holes, i.e., the white portion of region $(iii)$ in Fig.~\ref{f:stab-waq=1}, is larger than the preceding cases. Here it is bounded from below by the upper branch of the curve $c_1$ (for $1\geq\omega\gtrsim -1.7071$) and by $c_4$ (for $-\infty< \omega\lesssim -1.7071$), and it is bounded from above by the gray region, which extends from $\omega=1$ to $\omega\to-\infty$. 
Also, note that stable extreme regular black holes are found on the whole upper branch of the line $c_1$.

The stable portions of line $c_5$ and of region $(vi)$ are similar to the preceding cases for $1/\sqrt{3}< q/R< 1$.

\subsubsection{The case with $q/R=3\sqrt{3}/4$} 

\begin{figure}[h]
\centering 
\includegraphics[width=0.37\textwidth]{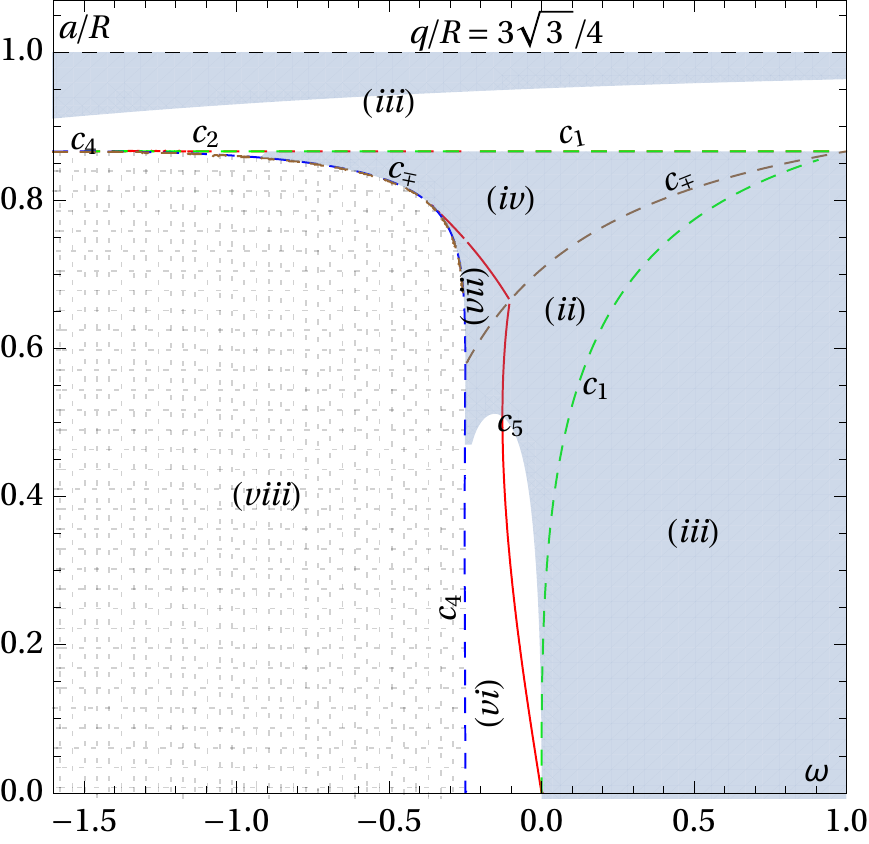}
\caption{Stability (white) and instability (light gray) regions for $q/R=3\sqrt{3}/4\simeq 1.3$ in the $(\omega,\, a/R)$--plane. 
The gridded region $(viii)$ contains no physical solutions.  }
\label{f:stab-waq=1.3} 
\end{figure} 
 
Figure~\ref{f:stab-waq=1.3} shows the results of the stability analysis for $q/R=3\sqrt{3}/4\simeq 1.3$ in the $(\omega,\, a/R)$--plane. 
This special case is chosen because it is on the boundary surface that separates  domains that contain a type $(ii)$ region above the curve $c_{\sst\mp}$, as in Figs.~\ref{f:stab-waq=02}--\ref{f:stab-waq=1}, from other domains that do not present such a region, cf. Figs~\ref{f:stab-waq=1.3}--\ref{f:stab-waq=10}. 
In this case, the line $c_2$ and the upper branch of $c_1$ coincide. Stable configurations are found in regions $(ii)$, $(iii)$ and $(vi)$.

A small part of the overcharged stars belonging to the region $(ii)$ are stable against small radial perturbation. The stable configurations are found for parameters in the intervals $-0.1283\lesssim\omega <0$ and $0< a/R\lesssim 0.5082$, similarly to the stable portion of the lower region $(ii)$ of the preceding case, cf. Fig.~\ref{f:stab-waq=1}. 

A large portion of region $(iii)$ contains stable regular charged black holes, the white region above curves $c_2 (c_1)$ and $c_4$ which extents form $\omega=1$ to $\omega \to-\infty$, becoming vanishingly slim as $\omega$ decreases to large negative values. 
In the present case,  it is not possible to verify the stability of the extreme regular black holes lying on the whole upper branch of the line $c_1$. These special black holes contain a massless thin shell, since the configurations belong also to curve $c_2$ and the criterion adopted here fails.

Overcharged stars with zero total mass $(m/R=0)$ are stable on the segment of line $c_5$ for  $-0.1283\lesssim\omega<0$.  The stable portion of region $(vi)$ are similar to the preceding cases for $0.78\leq q/R\leq 1$.

\subsubsection{The case with  $q/R=3/2$}

\begin{figure}[h]
\centering 
\includegraphics[width=0.37\textwidth]{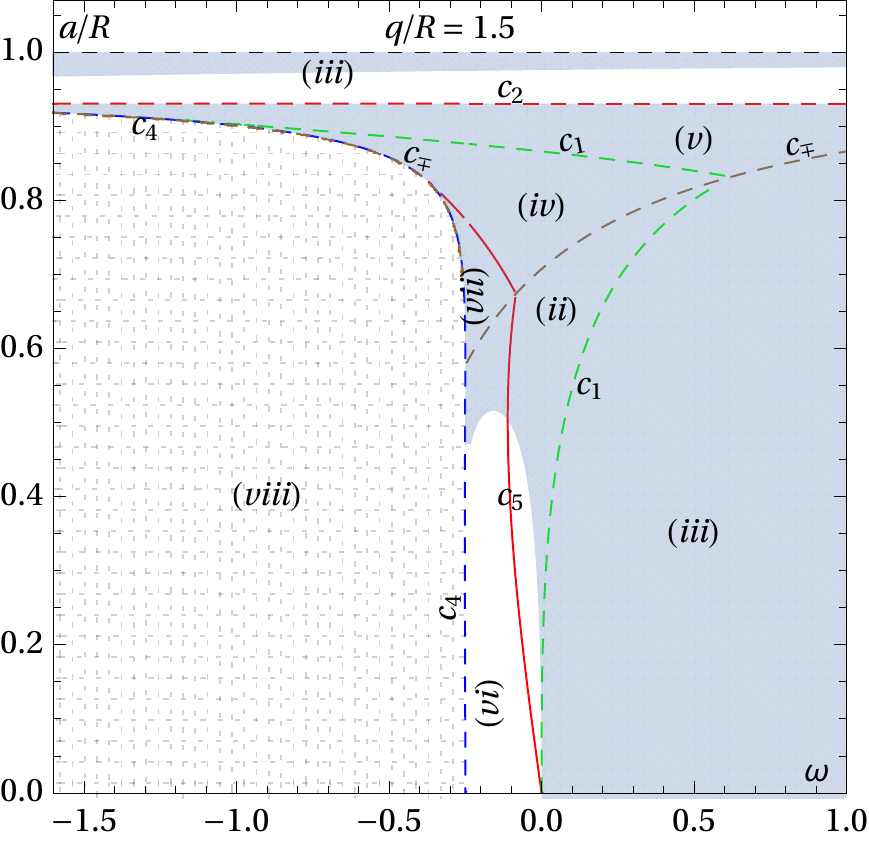}
\caption{Stability (white) and instability (light gray) regions for $q/R=3/2$ in the $(\omega,\, a/R)$--plane. 
The gridded region $(viii)$ contains no physical solutions.  }
\label{f:stab-waq=1.5} 
\end{figure}

Figure~\ref{f:stab-waq=1.5} shows the results of the stability analysis for $q/R=3/2$ in the $(\omega,\, a/R)$--plane. 
This case is chosen as  representative of all cases for electric charges in the interval $3\sqrt{3}\,/4<q/R< \sqrt{3}$. The curve $c_2$ separates the upper region $(iii)$ from region $(v)$ allowing the appearance of regular black holes without a thin shell at the boundary. Stable configurations are found in regions $(ii)$, $(iii)$, and $(vi)$.

A small portion of region $(ii)$ presents stable configurations representing overcharged stars. The range of parameters and properties of the solutions are similar to the case shown in Fig.~\ref{f:stab-waq=1.3}.

A significant portion of that region $(iii)$ bears stable regular black hole configurations, while all region $(v)$ represent unstable regular black holes. The stable region is a band just above the line $c_2$ whose width depends on $\omega$. 
For $\omega$ close to unity, the band width is from $a/R\simeq 0.9306$ to  $a/R \simeq 0.9796$, and it slowly shrinks while $\omega$ decreases, being from $a/R\simeq 0.9306$ to $a/R\simeq 0.9641$ for $\omega =-2$.
The criterion fails to by applied to configurations on $c_2$, since the derivative of the potential \eqref{eq:derivada2V} is not well defined there.

A big part of region $(vi)$ is also stable, but the configurations from that region are of little interest for carrying negative gravitational mass.

\subsubsection{The case with $q/R=\sqrt{3}$}

\begin{figure}[h]
\centering 
\includegraphics[width=0.37\textwidth]{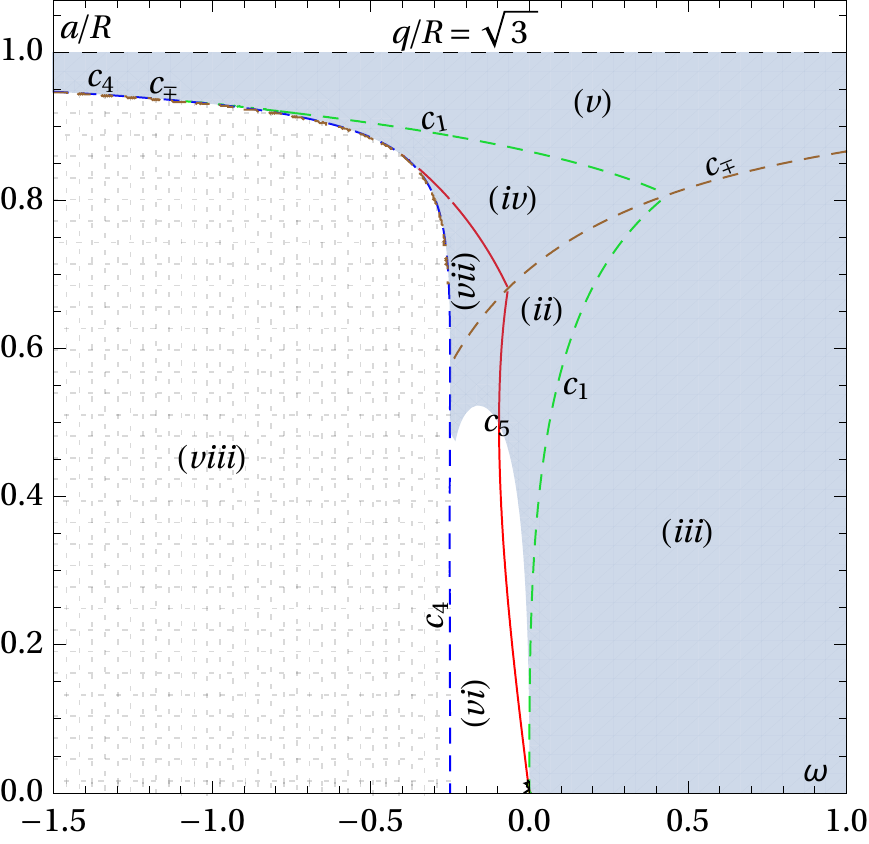}
\caption{Stability (white) and instability (light gray) regions for $q/R=\sqrt{3}$ in the $(\omega,\, a/R)$--plane. 
The gridded region $(viii)$ contains no physical solutions.}
\label{f:stab-waq=sqrt3} 
\end{figure}

Figure~\ref{f:stab-waq=sqrt3} shows the results of the stability analysis for $q/R=\sqrt{3}$ in the $(\omega,\, a/R)$--plane.  
This special case is chosen because it is on the boundary surface that separates domains that contain type $(iii)$ regions above the line $c_{\sst\mp}$, as in Figs.~\ref{f:stab-waq=02}--\ref{f:stab-waq=1.5}, from other domains that do not present such a region, cf. Figs~\ref{f:stab-waq=sqrt3} and \ref{f:stab-waq=10}, in the parameter space $(\omega,\, a/R,\, q/R)$.  In this case, the line $c_1$ presents just one branch, and line $c_2$, beside $c_3$ is not present. Stable configurations are found in regions $(ii)$ and $(vi)$ alone.

The white portion of region $(ii)$ is a slim strip between the curves $c_1$ and $c_5$. The configurations found there are stable overcharged stars with total gravitational mass varying from $m/R\sim 0$ (close to $c_5$) to $m/R\sim \sqrt{3}$ (close to $c_1$).

There are also stable solutions in region $(vi)$, but these are less interesting than the configurations of other regions since in region $(vi)$ the total mass is negative.

\subsubsection{The case with $q/R=10$}

\begin{figure}[h]
\centering 
\includegraphics[width=0.37\textwidth]{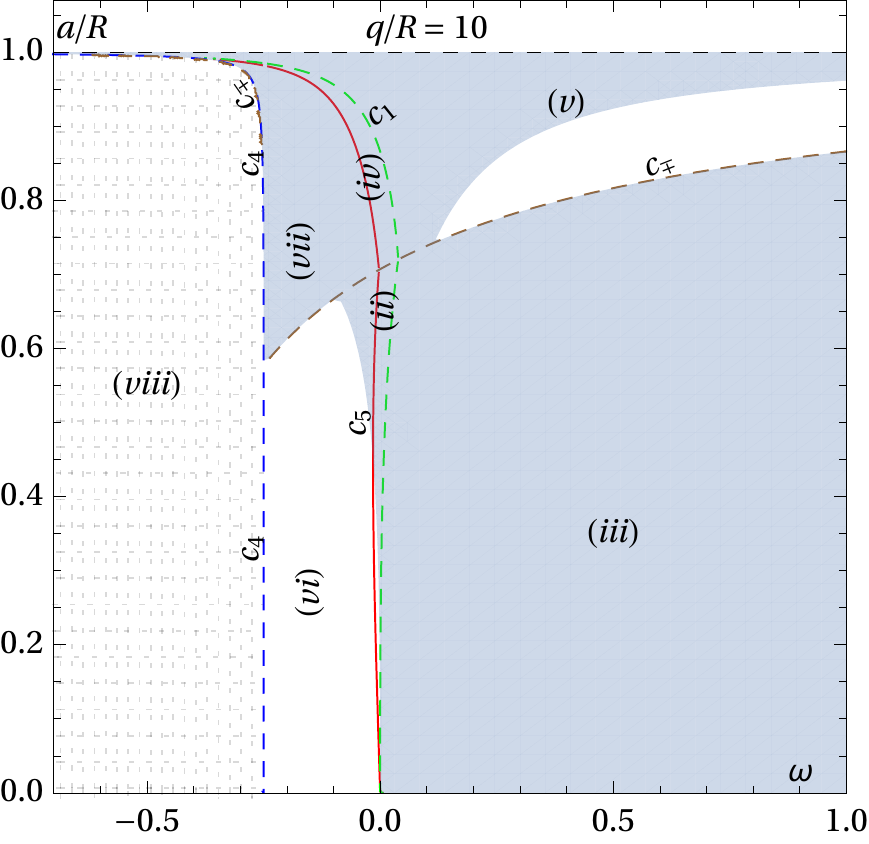}
\caption{Stability (white) and instability (light gray) regions for $q/R=10$ in the $(\omega,\, a/R)$--plane. 
The gridded region $(viii)$ contains no physical solutions. }
\label{f:stab-waq=10} 
\end{figure}

Figure~\ref{f:stab-waq=10} shows the results of the stability analysis for $q/R=10$ in the $(\omega,\, a/R)$--plane.  
This case is chosen for completeness, in order to show the general behavior of solutions for high values of electric charge. Here, the stable configurations are found in three regions $(ii)$, $(v)$, and $(vi)$.

Region $(ii)$ is slim in this case, it tends to disappear for large values of $q/R$, with curves $c_1$ and $c_5$ tending to coincide. It presents a very tiny stable portion for the lower values of $a/R$ and $\omega$ close to zero, namely, for $0 < a/R\lesssim 0.4493$ and $-0.01537\lesssim \omega\leq 0$, hardly seen in the figure. 

The stable (white) portion of the region $(v)$ located above the curve $c_{\sst\mp}$ appears in the case $q/R\simeq 2.6224$ and grows with $q/R$. This region contains regular charged black holes with a thin shell of negative mass, whose gravitational mass is quite larger than the electric charge.

A major part of region $(vi)$ bears stable configurations representing object with negative gravitational mass. Region $(vi)$ increases as $q/R$ grows, and so does the corresponding stable (white) portion of it.

\subsection{Regions of stability in the \boldmath{$(q/R, a/R)$}--plane}

\subsubsection{General remarks}

In the present section we show the results of the stability analysis in the $\left(q/R,\,a_{0}/R\right)$--plane, by choosing some fixed values of the parameter $\omega$. As above, the parameters $a_0$ and $q$ are normalized with respect to $R$. Once again, to simplify notation, we drop the index ``$0$'' of the symbols denoting equilibrium quantities, $a_0\rightarrow a$, $M_0\rightarrow M$, etc.
The results are given in terms of a series of graphs presented in Figs.~\ref{f:stab-qaw=1}--\ref{f:stab-qaw=-1}. The notation and conventions in drawing such graphs are the same as the ones employed in the preceding section.

\subsubsection{The case with $\omega=1$}

\begin{figure}[h]
\centering 
\includegraphics[width=0.37\textwidth]{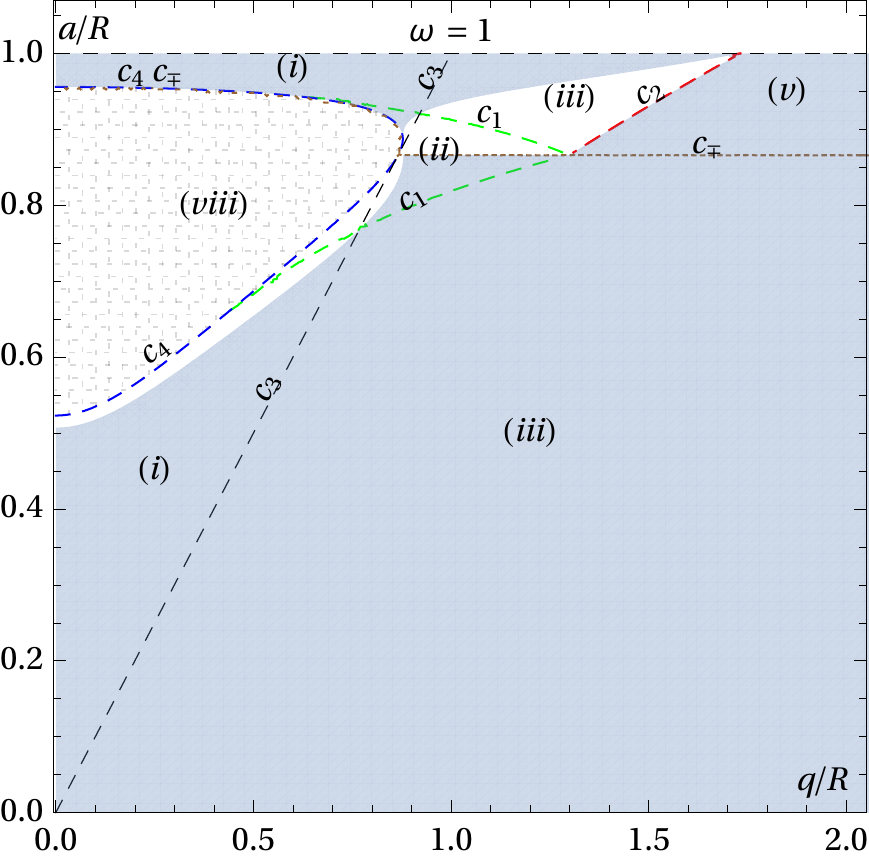}
\caption{Stability (white) and instability (light gray) regions for $\omega=1$ in the  $(q/R,\, a/R)$--plane.
The gridded region $(viii)$ contains no physical solutions.}
\label{f:stab-qaw=1} 
\end{figure}

This model is characterized by a thin shell containing stiff matter, which is represented by the corresponding equation of state in the form $\mathcal{P}=\sigma$. This case has been chosen as a representative situation of all cases with $1/2\leq \omega \leq 1$. The main results of the stability analysis for this case are shown in Fig.~\ref{f:stab-qaw=1}. As in the cases analyzed in the previous section, the white regions represent stable solutions and the light gray regions contains the unstable solutions but now in the $(q/R,\,a/R)$--plane. The gridded region $(viii)$ presents no real solutions.
According to the figure, stable solutions are found in the regions $(i)$, $(ii)$, and $(iii)$.

Stable undercharged stars (charged gravastars) are found in the region $(i)$ located below the curves $c_1$ and $c_4$. The white region is strip close to the line $c_4$, bounded from above by the curves $c_4$ and $c_1$. The transverse boundary on the left hand side of the strip is at $q/R=0$ and it extends up to $q/R\simeq 0.7681$, where $c_1$ intercepts $c_3$. The range of the radius of the shell along the strip is 
$0.5072\lesssim a/R\lesssim 0.7681$. The range of masses in the stable region is $0.2092\lesssim m/R\lesssim 0.7681$, where the lower limit occurs at the point ($q/R=0$, \,$a/R\simeq 0.5072$) and the upper limit corresponds to the point ($q/R\simeq 0.7681$,\, $a/R\simeq 0.7681$). 
 
Almost all the part of region $(ii)$ located above the line $c_{\sst\mp}$, and only a small portion of such a region below the lines $c_4$ and $c_{\sst\mp}$ bears stable configurations. This is the region of regular overcharged stars. The range of the masses of these stable configurations is $0.4410 \lesssim m/R < 3\sqrt{3}/4$.

Note that the configurations given by the segment of the curve $c_1$ (lower branch) bounded by the lines $c_3$ and $c_4$ are stable extremely charged stars with $m/R=q/R$. The relevant segment starts at the point $(q/R\simeq 0.4410,\, a/R\simeq 0.6614)$, on the curve $c_4$, and extends to the point  $(q/R\simeq 0.7681,\, a/R\simeq 0.7681)$, on the curve $c_3$, so that the range of masses of the stars on this segment is  $0.4410\lesssim m/R\lesssim 0.7681$.

The configurations represented by the segment of the line $c_3$ inside the region $(ii)$ are stable extreme quasiblack holes. The  masses of these stable configurations are in the range $0.7681\lesssim m/R\lesssim 0.9207$, which is the same for $q/R$ and $a/R$, since that on this case one has $q/R=a/R=m/R$

A large portion of region $(iii)$ for $q/R$ and $a/R$ in the ranges $0.9207 \lesssim q/R < \sqrt{3}$ and $\sqrt{3}/2 < a/R < 1$, respectively, contains stable regular black holes with masses in the range $0.9207\lesssim m/R < 2.000$.

Moreover, stable extremely charged ($m/R=q/R$) regular black holes are found on the segment of the line $c_1$ (the upper branch, located above the line $c_{\sst\mp}$) that is at the boundary between regions $(ii)$ and $(iii)$, for $0.9207\lesssim q/R < 3\sqrt{3}/4$.

\subsubsection{The case with $\omega=0.15$} 

\begin{figure}[h]
\centering 
\includegraphics[width=0.37\textwidth]{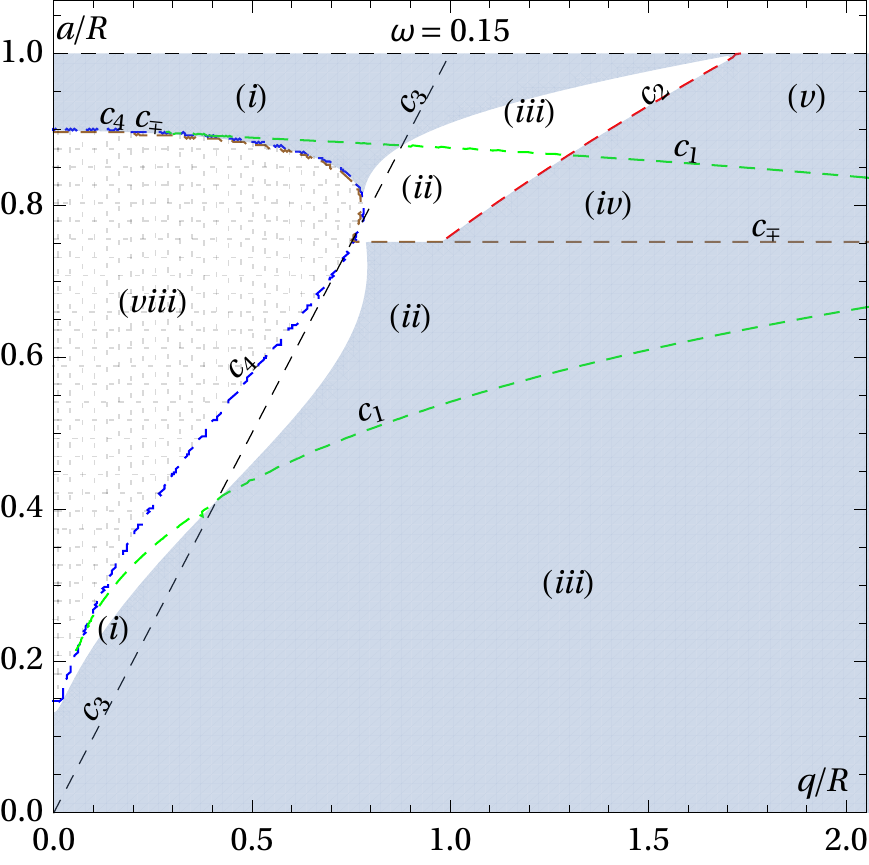}
\caption{Stability (white) and instability (light gray) regions for $\omega=0.15$ in the  $(q/R,\, a/R)$--plane.
The gridded region $(viii)$ contains no physical solutions.}
\label{f:stab-qaw=015} 
\end{figure}

This model is characterized by a thin shell containing a perfect fluid with pressure $\mathcal{P}=0.15\sigma$. This  case is chosen as a representative situation of all cases for $\omega$ in the interval $0<\omega< 1/2$, which show six of the eight different regions in the parameter space as described in Sec.~\ref{sec:regions}, five of them being of interest.  The results of the stability analysis are shown in Fig.~\ref{f:stab-qaw=015}, where white regions contain stable solutions and the light gray regions contain unstable solutions in the $(q/R,\,a/R)$--plane. As in the case for $\omega=1$, the stable solutions are found in the regions  $(i)$, $(ii)$ and $(iii)$.

Region $(i)$ shows stable solutions just in a slim area of the branch located below the line $c_{\sst\mp}$,  between the lines $c_1$ and $c_3$. The parameters of these stable charged gravastars are in the intervals $0.1322\lesssim a/R\lesssim 0.4109$ and $0<q/R\lesssim 0.4109$, and represent charged stars with masses in the range $0.02723\lesssim m/R\lesssim 0.4109$. 

A large part of region $(ii)$ presents stable regular overcharged stars. Such stable solutions are found in both branches of region $(ii)$ located below and above the line $c_{\sst\mp}$.
Since these stability regions are in the vicinity of the curves $c_1$ and $c_3$, the masses of the solutions are close to the electric charge values, being approximately in the range 
$0.04687\lesssim m/R < 3\sqrt{3}/4$, with  $q/R$  approximately in the same interval, and with radius in the range $0.2031\lesssim a/R \lesssim 0.8790$.

The configuration represented by the line $c_3$ lying inside the region $(ii)$ are all stable extreme quasiblack holes. The masses of these configurations are in the range $0.4109\lesssim m/R\lesssim 0.8790$.

As in the case with $\omega=1$, the white portion of region $(iii)$ is the lower part, close to the curves $c_1$ and $c_2$. It contains charged regular black holes that are stable against radial perturbation of the thin shell. 
The charges and thin shells radii of such objects are in the intervals $0.8790\lesssim q/R < \sqrt{3}$ and $\sqrt{3}/2 < a /R < 1$, respectively, while the masses are in  the range $0.8790\lesssim m/R < 2.000$.

The extreme regular black holes found on the segment of line $c_1$ for $0.8790\lesssim q/R=m/R < 3\sqrt{3}/4$, i.e., the segment of $c_1$ between regions $(ii)$ and $(iii)$, are also stable solutions.

\subsubsection{The case with $\omega=0$} 

\begin{figure}[h]
\centering 
\includegraphics[width=0.37\textwidth]{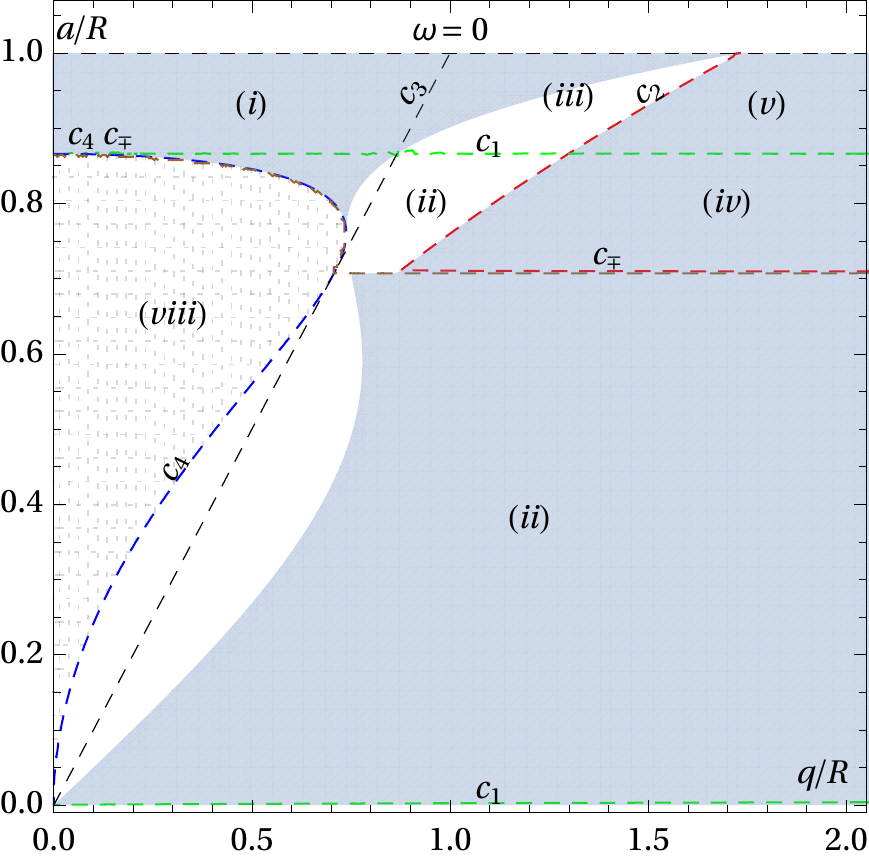}
\caption{Stability (white) and instability (light gray) regions for $\omega=0$ in the  $(q/R,\, a/R)$--plane.
The gridded region $(viii)$ contains no physical solutions. }
\label{f:stab-qaw=0} 
\end{figure}

This model is characterized by a thin shell containing a fluid of zero pressure $\mathcal{P}=0$, i.e., it represents a thin shell of dark matter. This is a special case and thus deserves a separate study. The results of the stability analysis are shown in Fig.~\ref{f:stab-qaw=0}. Here, stable solutions are found just in the regions $(ii)$ and $(iii)$.

The large white portion of region $(ii)$ below the line $c_{\sst\mp}$ and close to curve $c_4$ bears stable overcharged stars. The stable region extends from the origin, at $q/R\simeq 0$, $a/R\simeq 0$ (also with $m/R\simeq 0$), to the point where the lines $c_3$, $c_4$, and $c_{\sst\mp}$ meet all together. That is the point  $q/R=a/R=\sqrt{2}/2$, and also with $m/R=\sqrt{2}/2$.
Additionally, a major portion of region $(ii)$ located above the line $c_{\sst\mp}$ contains stable configurations. The range of parameters of the stable overcharged stars in this region are similar to the case of Fig.~\ref{f:stab-qaw=015}, for $\omega=0.15$.

The whole segment of line $c_3$ inside the region $(ii)$ contains stable quasiblack holes. Such a segment starts at $q/R=0=a/R$ and extends to the point $q/R=\sqrt{3}/2=a/R$, excluding the endpoints. The range of masses of the corresponding stable configurations is the same, $0< m/R<\sqrt{3}/2$.

As in the cases with $\omega=1$ and $\omega=0.15$, the white portion of region $(iii)$ is the lower part, close to the curves $c_1$ and $c_2$. It contains charged regular black holes that are stable against radial perturbation of the thin shell.

The extreme regular black holes found on the segment of line $c_1$ for $\sqrt{3}/2< q/R < 3\sqrt{3}/4$ are also stable solutions.

\subsubsection{The case with $\omega =-0.22$}

\begin{figure}[h]
\centering 
\includegraphics[width=0.37\textwidth]{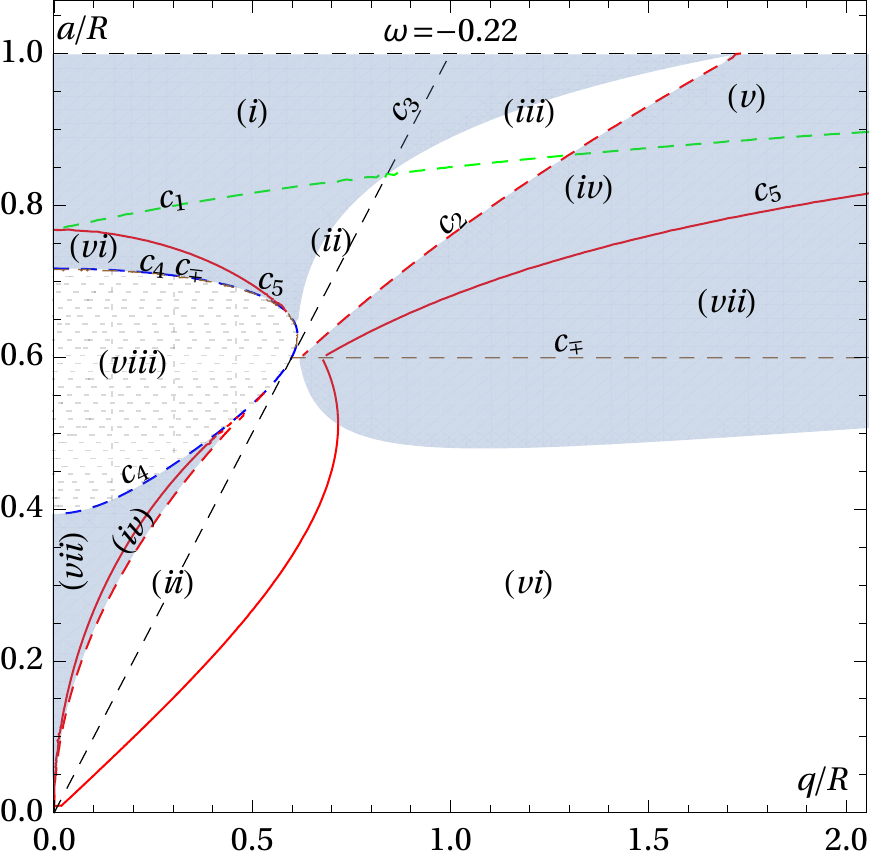}
\caption{Stability (white) and instability (light gray) regions for $\omega=-0.22$ in the  $(q/R,\, a/R)$--plane.
The gridded region $(viii)$ contains no physical solutions.}
\label{f:stab-qaw=-022} 
\end{figure}

This first case of negative pressure, for which the equation of state is $\mathcal{P}=-0.22\sigma$, is chosen as being representative of all cases with $\omega$ in the interval $-1/4 < \omega<0$, whose matter on the thin shell may be interpreted as some kind of dark energy, or representing a tension shell. The key features in the diagram for this case is the existence of two branches of the curve for zero gravitational mass that meet each other at the point $(q/R=0,\, a/R=0)$, and the existence of a pair of each one of the regions $(ii)$, $(iv)$, $(vi)$, and $(vii)$.
The results of the stability analysis in the $(q/R,\,a/R)$--plane are shown in Fig.~\ref{f:stab-qaw=-022}, where the conventions are the same as the preceding figures.  Here, stable solutions are found in the regions $(ii)$, $(iii)$, and $(vi)$.

Stable overcharged stars are found in the white portions of region $(ii)$. The ranges and sizes of these portions are very similar to the case for $\omega=0$ shown in Fig.~\ref{f:stab-qaw=0}, and then we do not comment further on this here.  

Stable quasiblack hole configurations found on the segment of line $c_3$ that is inside region $(ii)$. It starts at the point $q/R=a/R=0$ and extends till the point where line $c_3$ meets the line $c_1$, at $q/R=a/R\simeq 0.8404$. The masses of these quasiblack holes are in the same range as the electric charge.

As in the previous cases, the white portion of region $(iii)$ is the lower part, close to the curves $c_1$ and $c_2$. 
It contains charged regular black holes that are stable against radial perturbation of the thin shell. The range of masses of these configurations is $0.8404\lesssim m/R < 2.000$, for electric charges and thin shell radii respectively in the intervals $0.8404\lesssim q/R < \sqrt{3}$ and  $0.8404\lesssim a/R < 1$.

The extreme regular black holes found on the segment of line $c_1$ for $0.8404 \lesssim q/R < 3\sqrt{3}/4$ are also stable solutions. 

A large portion of region $(vi)$ also present stable configurations. This stability may be understood by taking into account the negative (repulsive) gravitational mass of the object, that sustains the massive shell.

\subsubsection{The case with $\omega=-0.25$}

\begin{figure}[h]
\centering 
\includegraphics[width=0.37\textwidth]{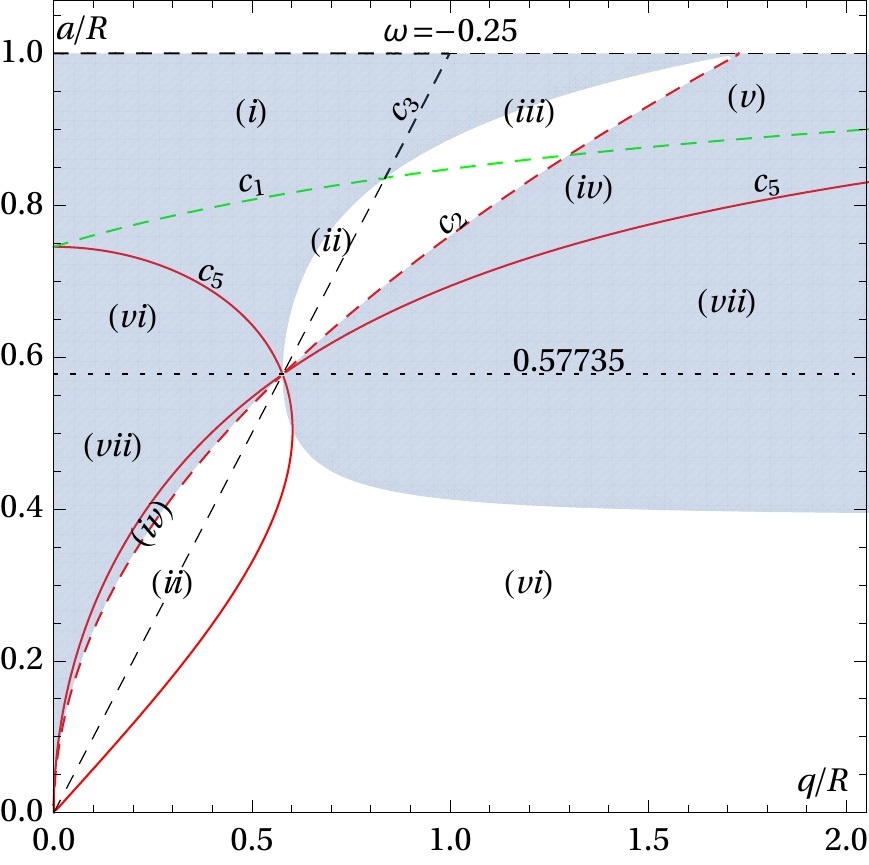}
\caption{Stability (white) and instability (light gray) regions for $\omega=-0.25$ in the  $(q/R,\, a/R)$--plane. The dotted line $a/R=1/\sqrt{3}\simeq 0.57735$ represents singular solutions.}
\label{f:stab-qaw=-025} 
\end{figure}

This very special case also deserves a separate study, and for that we must employ the mass functions given by Eqs.~\eqref{eq:totalmass} and \eqref{eq:shellmassw1o4}. The thin shell is made up by a fluid of negative pressure $\mathcal{P}=-0.25\,\sigma$, which may also be interpreted as a tension shell. The results of the stability analysis in the $(q/R,\,a/R)$--plane are shown in Fig.~\ref{f:stab-qaw=-025}.
The special features of the corresponding diagram is the absence of region $(viii)$, and the existence of a meeting point $q/R=a/R=1/\sqrt{3}$, where all relevant lines except $c_1$ converge to. 
The dotted line $a/R=1/\sqrt{3}\simeq 0.57735$ represent singular solutions for any value of electric charge $q/R$.
 As in the case of Fig.~\ref{f:stab-qaw=-022}, here stable solutions are found in the regions $(ii)$, $(iii)$, and $(vi)$. 

All the large branch of region $(ii)$ located below the dotted line at $a/R=1/\sqrt{3}$, and a significant part of that region above such a line, represent stable configurations. 
They are regular overcharged stars whose masses varies in the range $0 < m/R < 3\sqrt{3}/4 $, with the largest values of masses coming from configurations close to the intersection between the lines $c_1$ and $c_2$.

A portion of region $(iii)$ for $q/R$ and $a/R$ in the ranges $\sqrt{1+2\sqrt{7}}/3\simeq 0.8361\lesssim q/R<\sqrt{3}$ and $ 0.8361\lesssim a/R<1$ contains stable regular charged black holes. The masses of these objects are in the interval $ 0.8361 \lesssim m/R< 2.000$.
  
Stable extremely charged black holes appear on the segment of line $c_1$ located between regions $(ii)$ and $(iii)$, with electric charges (and masses) in the intervals $0.8361\lesssim q/R=m/R < 3\sqrt{3}/4$.
 
A large portion of region $(vi)$ also presents stable configurations with negative total gravitational masses.

\subsubsection{The case with $\omega=-0.27$}

\begin{figure}[h]
\centering 
\includegraphics[width=0.37\textwidth]{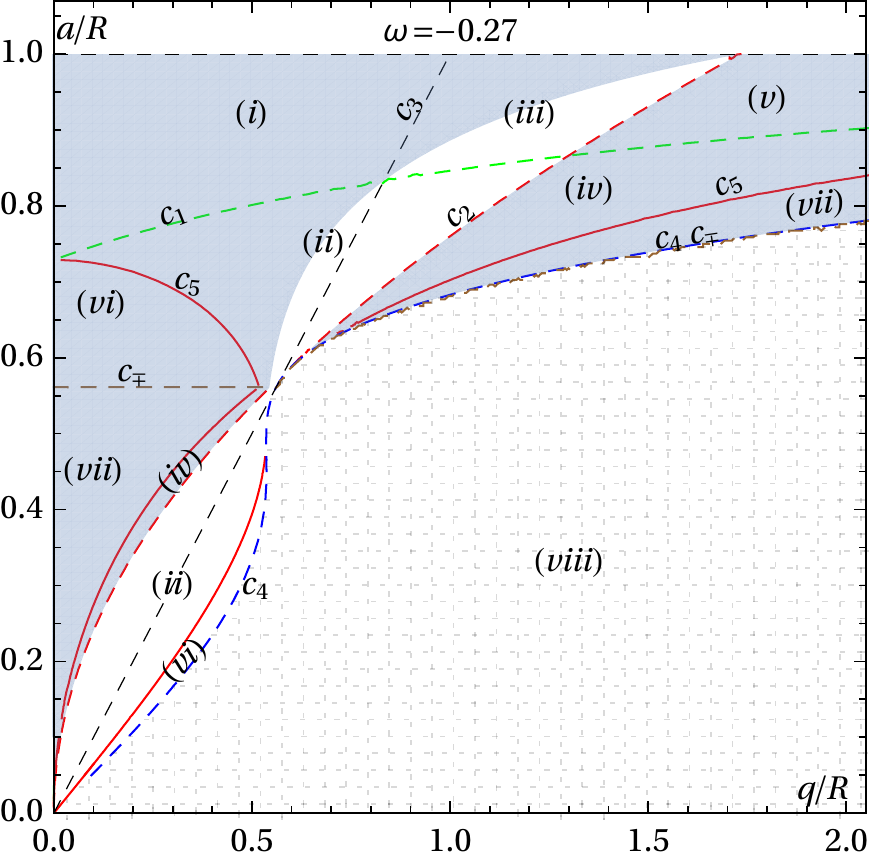}
\caption{Stability (white) and instability (light gray) regions for $\omega=-0.27$ in the  $(q/R,\, a/R)$--plane.
The gridded region $(viii)$ contains no physical solutions.}
\label{f:stab-qaw=-027} 
\end{figure}

This model is characterized by a thin shell containing a fluid of negative pressure $\mathcal{P}=-0.27\,\sigma$, which may be interpreted as a tension shell or as some kind of dark energy. 
This case is chosen because it is representative of all situations for $\omega$ in the range $-0.40\lesssim \omega< -1/4$.
The results of the stability analysis are shown in Fig.~\ref{f:stab-qaw=-027}. The special new feature in comparison to the previous cases for $\omega > -1/4$ is the presence of a large region $(viii)$ on the bottom right corner of the diagrams. As in the previous cases for $\omega < 0 $, the stable solutions are found just in the regions $(ii)$, $(iii)$, and $(vi)$. 

Stable regular overcharged stars are found in a significant (white) portion of region $(ii)$.

Stable quasiblack holes are found on the segment of the line $c_3$ located inside region $(ii)$.

Stable regular black holes are found in the lower (white) portion of region $(iii)$, close to the curves $c_1$ and $c_2$. 

Stable extreme regular black holes are found on the segment of the line $c_1$ that separates the white parts of regions $(ii)$ and $(iii)$.

The range of masses, charges and radius of the thin shell are similar to the cases for $\omega=-0.22$ and $\omega =-0.25$. 
For instance, the masses of the stable regular black holes of region $(iii)$ are in the range $0.8330\lesssim m/R <2.000$, with charges and radii respectively in the intervals $0.8330\lesssim q/R < \sqrt{3}$ and  $0.8330 \lesssim a/R<1$.

The branch of region $(vi)$ located between the curves $c_4$ and $c_5$ contains stable configurations with negative total mass.

\subsubsection{The case with $\omega=-0.40$}

\begin{figure}[h]
\centering 
\includegraphics[width=0.37\textwidth]{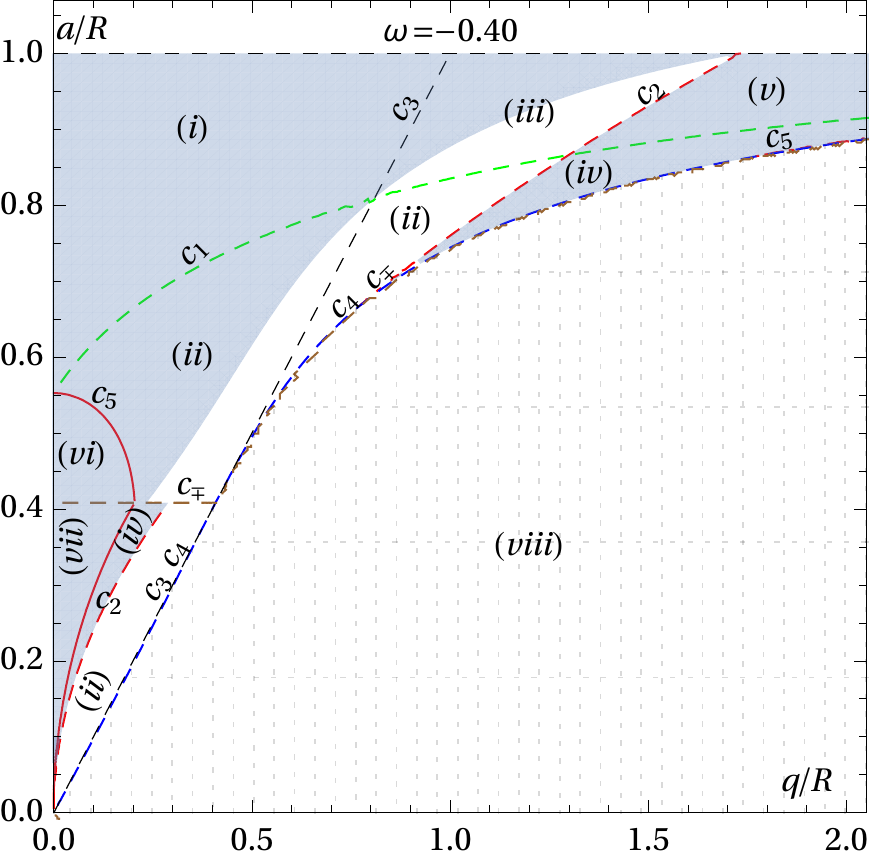}
\caption{Stability (white) and instability (light gray) regions for $\omega=-0.40$ in the  $(q/R,\, a/R)$--plane.
The gridded region $(viii)$ contains no physical solutions.}
\label{f:stab-qaw=-040} 
\end{figure}

This model is characterized by a thin shell containing a fluid of negative pressure, $\mathcal{P}=-0.40\,\sigma$.
This case is chosen because it is representative of all situations for which $-1/2< \omega \leq -0.40$.
The key feature in regard to the preceding cases with negative $\omega$ is that only one of the branches of the line $c_5$ extends to the point $q/R=a/R=0$, and then the branch of region $(vi)$ located below the curve $c_{\sst\mp}$, on the right of the curve $c_3$, is not present.
Moreover, the lower part of the line $c_3$ coincides with the line $c_4$.
The results of the stability analysis are shown in Fig.~\ref{f:stab-qaw=-040}.  
Here, the stable solutions are found just in regions $(ii)$ and $(iii)$. 

Stable regular overcharged stars are found in a significant (white) portion of region $(ii)$. The range of masses of these solutions is $0 < m/R < 3\sqrt{3}/4$, the lower limit corresponding to the zero charge case.

As in the previous cases, the configurations on the segment of line $c_3$ located inside the region $(ii)$, whose masses vary in the interval $0< m/R \lesssim 0.8097$, are stable quasiblack holes. These solutions are stable even in the part of $c_3$ that coincides with the line $c_4$.

Stable regular black holes are found in the lower (white) portion of region $(iii)$, close to the curves $c_1$ and $c_2$. The range of masses of these stable solutions are similar to the preceding cases with negative $\omega$, namely, $0.8097\lesssim m/R < 2.000$.

The extreme regular black holes on the segment of line $c_1$ that separates region $(ii)$ from region $(iii)$, with charges and masses in the interval $0.8097\lesssim q/R=m/R < 3\sqrt{3}/4$, are also stable solutions.

\subsubsection{The case with $\omega=-1/2$}

\begin{figure}[h]
\centering 
\includegraphics[width=0.37\textwidth]{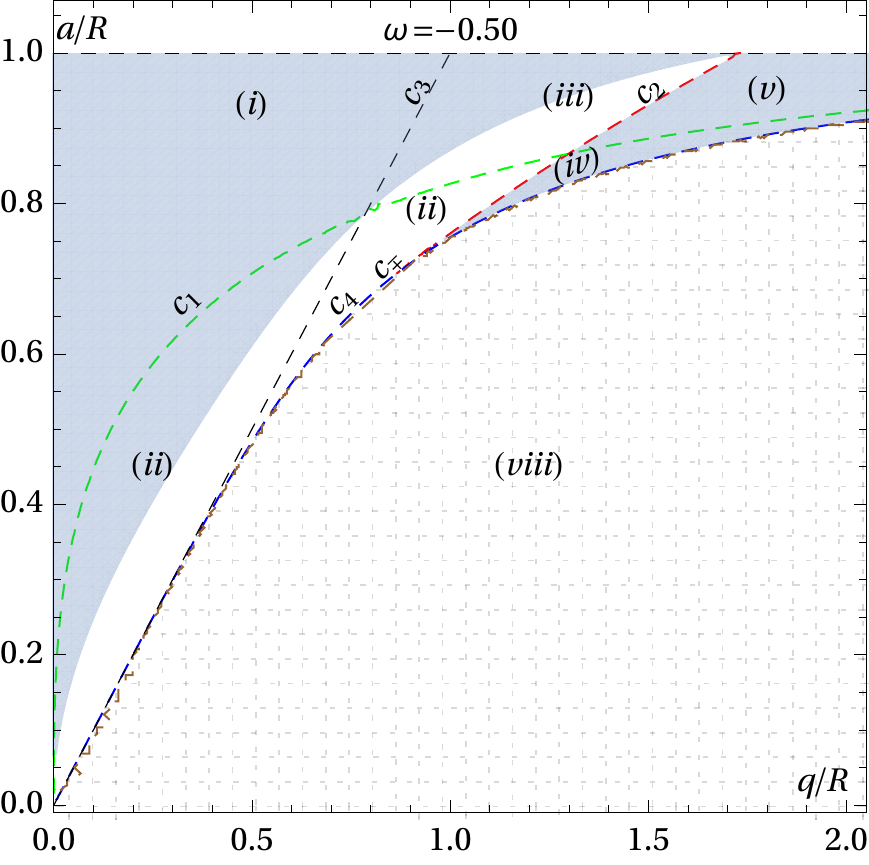}
\caption{Stability (white) and instability (light gray) regions for $\omega=-1/2$ in the  $(q/R,\, a/R)$--plane.
The gridded region $(viii)$ contains no physical solutions.}
\label{f:stab-qaw=-05} 
\end{figure}

The state equation is for the fluid in the shell is $\mathcal{P}=-\sigma/2$. 
The results of the stability analysis in the $(q/R,\,a/R)$--plane are shown in Fig.~\ref{f:stab-qaw=-05}.  The particular features with respect to the preceding cases with negative $\omega$ are that line $c_1$ extends down to the origin $q/R=a/R=0$, the lower branch of line $c_2$, the line $c_5$, and the left branch of $c_{\sst\mp}$ do not appear. As a consequence, the left branches of regions $(iv)$, $(vi)$, and $(vii)$ close to the vertical axis $q/R=0$, and that are present in the cases for $0 < \omega < -1/2$, also disappear, see  Figs.~\ref{f:stab-qaw=-022}--\ref{f:stab-qaw=-040}. 
As in the case of Fig.~\ref{f:stab-qaw=-040}, stable solutions are found just in the regions  $(ii)$ and $(iii)$. 

The white portion of region $(ii)$ is close to the curve $c_4$ and $c_3$, it presents stable overcharged stars with masses in the range $0< m/R <3\sqrt{3}/4$, while the charges and the radii of the spheres vary in the same intervals, i.e., $0< q/R <3\sqrt{3}/4$ and  $0< a/R <\sqrt{3}/2$.

The configurations on the segment of line $c_3$ located inside the region $(ii)$ are stable quasiblack holes. These solutions are stable even in the part of $c_3$ that coincides with the line $c_4$ and the masses, charges and radius vary in the interval $0 <m/R=q/R=a/R\lesssim 0.7862$.

The region of stable regular black holes, the lower (white) portion of region $(iii)$, close to the curves $c_1$ and $c_2$ is a little larger than in the preceding cases with negative $\omega$. The masses of these stable configurations are in the range $ 0.7862\lesssim m/R< 2.000$.

The extreme regular black holes found on the segment of line $c_1$ that separates regions $(ii)$ and $(iii)$, for $0.7862\lesssim q/R<3\sqrt{3}/4$, are also stable solutions.

\subsubsection{The case with $\omega=-1$}

In this very special and interesting case the state equation for the fluid on the shell is $P=-\sigma$, similar to the cosmological constant term. Equation \eqref{eq:shell-density} implies in $\sigma^{\prime}=0$,  so that the energy density $\sigma$ and the pressure of the junction surface are both constant parameters, independent of the radial size of the shell. This means that all configurations for $\omega=-1$ present a thin shell with the same energy density and pressure (tension).

\begin{figure}[h]
\centering 
\includegraphics[width=0.37\textwidth]{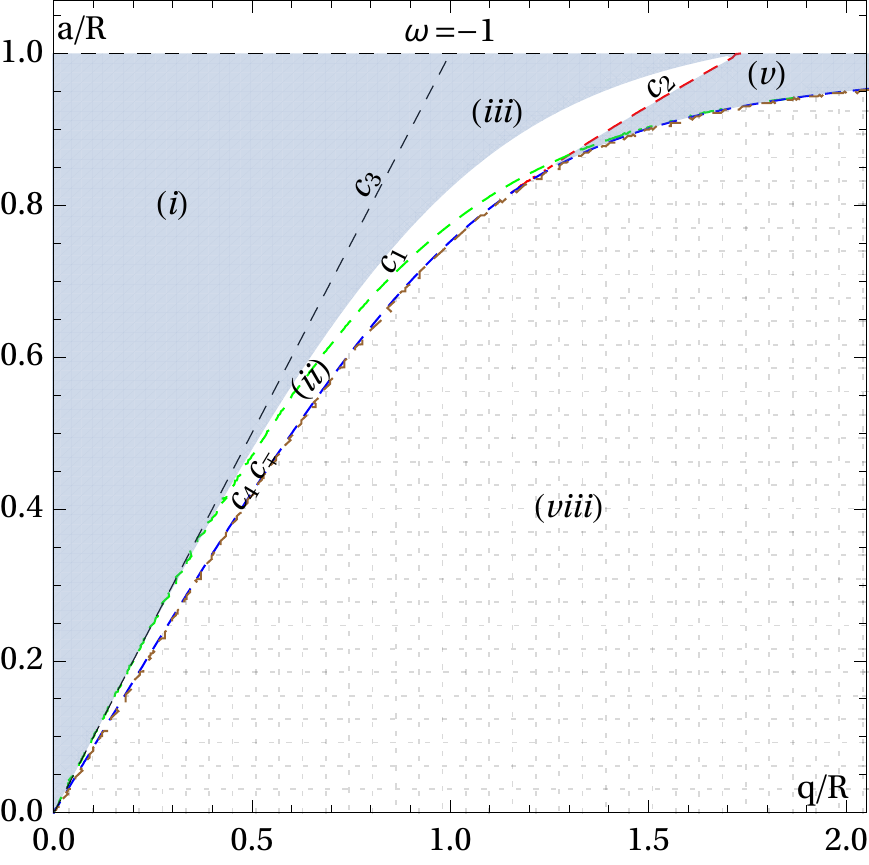}
\caption{Stability (white) and instability (light gray) regions for $\omega=-1$ in the  $(q/R,\, a/R)$--plane.
The gridded regions $(viii)$ contains no physical solutions.}
\label{f:stab-qaw=-1} 
\end{figure}

The results of the stability analysis  in the $(q/R,\,a/R)$--plane are summarized in Fig.~\ref{f:stab-qaw=-1}. In this case, the lower part of the line $c_3$ does not coincide with line $c_4$, and, moreover, the whole line $c_1$ is on the right-hand side of the line $c_3$. This implies that the region $(ii)$ is now bounded by the lines $c_1$ and $c_4$. 
It is seen from the figure, stable solutions are found just in the regions $(ii)$ and $(iii)$.

 The whole region $(ii)$ shows stable regular overcharged stars whose masses are in the interval $0 < m/R < 3\sqrt{3}/4$.

As in the previous cases, the stable part of region $(iii)$ is close the curves $c_1$ and $c_2$. Region $(iii)$ extends to the origin of the diagram $q/R=a/R=0$, becoming very slim as $q/R$ and $a/R$ tend to zero, with the stable portion always present, coasting the line $c_1$ till the origin.  
The upper limit of masses, charge and radius are the same as all figures in the $(q/R,\, a/R)$--plane, see Figs.~\ref{f:stab-qaw=1}--\ref{f:stab-qaw=-05}, while the lower limit is null, i.e., the masses of the stable regular black holes are in the range $0<m/R< 2.000$.
 
The whole segment of the line $c_1$ from the origin to the point $(q/R=3\sqrt{3}/4, \, a/R= \sqrt{3}/2$, that is in the boundary of regions $(ii)$ and $(iii)$,  bears stable extreme regular black holes.

\section{Conclusion}\label{sec:conclusion}

Let us mention once again that the present work is the continuation of the previous work of Ref.~\cite{Masa:2018elb}, where new models for charged spherically symmetric compact objects with a thin shell of matter at the boundary were presented. 
The matter inside the shell is a nonisotropic fluid satisfying a de Sitter equation of state of the form $ 8\pi 
p_{r}= -8\pi \rho_m =-\dfrac{3}{R^{2}}+\dfrac{q^{2}}{a^{4}}\left(\dfrac{r}{a
}\right)^{2(n+1)}$, where $R$, $q$, and $n$ are constant parameters, with $n\geq 0$, and $r$ being the areal radius coordinate. The solutions are given in terms of three parameters, namely, the normalized electric charge $q/R$, with $q$ being the total electric charge, the radius of the objects $a/R$, where $a$ coincides with the radius of the boundary shell, and a parameter $\omega$ introduced by means of a linear equation of state for the matter contained by the shell. Parameter $\omega$ is allowed to assume also negative values, representing some kind of dark energy. To avoid violation of causality, we restrict $\omega$ not to be larger than unity in units such that the speed of light is unity.

In the present work we have first completed the analysis of the solutions found in Ref.~\cite{Masa:2018elb} by studying further properties and exploring other regions of the parameter space. Here we investigate all kinds of equilibrium objects, i.e., with constant radius $a$, represented by the given solutions as a function of $\omega$ and $a/R$, by considering some fixed values of $q/R\geq0$. In the previous work the analysis was done in the $(q/R,\, a/R)$--plane. With the present analysis it is possible to see more clearly the properties of the compact objects as a function of the thin shell matter composition.  
 For all values of $ \omega$ investigated here, very interesting solutions, such as regular black holes, regular charged stars, quasiblack holes, charged gravastars, and regular overcharged stars are found in specific regions of the parameter space. 

In the sequence of the work we investigate the stability of the solutions against perturbations in the position of the shell by following the work of Ref.~\cite{uyf2012}.
For a better visualization of the results, first a detailed analysis of the stability and instability regions is performed in the $(\omega,\, a/R)$--plane of the parameter space, and the results are shown in a number of figures for several fixed values of the electric charge.  At the end, the stability analysis is performed in the $(q/R,\, a/R)$--plane by choosing several fixed values of the parameter $\omega$. 
The results show stable objects of all kinds in some regions of the parameter space. In particular, stable regular black holes, stable gravastars, stable quasiblack holes, and stable overcharged stars show up in large regions of the parameter space. 

The stable objects presented here may be generalized to more realistic situations, e.g., by including rotation, where the models may be compared to astrophysical objects. The fact that the present results by Ligo and first EHT observations do not exclude ultracompact objects as regular black holes is a good motivation for that study. 

Finally, let us point out that the matching conditions used in the present case apply just to shells following timelike trajectories, not allowing the analysis of lightlike shells, and neither the transition from a timelike to a lightlike trajectory. This last aspect may be of interest, in particular, in the study of the stability of gravastars, whose boundary is on the verge of being a lightlike surface and a small perturbation could lead to such a transition \cite{Reviewer}.
Due to the matching conditions, this kind of process is not allowed in the models we investigate here, but the subject is interesting and shall be considered in our future studies.

\section*{Acknowledgments}
A. D. D. M. was financed in part by Coordenação de Aperfeiçoamento
de Pessoal de Nível Superior (CAPES), Brazil, Finance Code 001.
E. S. O. was financed in part by Funda\c{c}\~ao de Amparo \`a Pesquisa do Estado de S\~ao Paulo (FAPESP, Brazil), Grant No.  2015/26858-7. 
V. T. Z. was partly financed by CAPES, Brazil, Grant No. 8881.310352/2018-01, and Conselho Nacional de Desenvolvimento Científico e Tecnológico (CNPq), Brazil, Grant No. 309609/2018-6.


\begin{thebibliography}{100}
 
\bibitem{Hawking:1974sw} S.~W.~Hawking,
Particle creation by black holes,
Commun. Math. Phys. \textbf{43}, 199 (1975).

\bibitem{Bekenstein:1973ur} J.~D.~Bekenstein,
Black holes and entropy,
Phys. Rev. D \textbf{7}, 2333 (1973).

 \bibitem{Ligo2016} B. P. Abbott et al. (LIGO and Virgo Collaborations),
Observation of gravitational waves from a binary black hole merger, Phys. Rev. Lett. \textbf{116}, 061102 (2016), 
\href{https://arxiv.org/abs/1602.03837}{arXiv:1602.03837 [gr-qc]}.

\bibitem{Akiyama:2019cqa} 
  K.~Akiyama {\it et al.} (Event Horizon Telescope Collaboration),
 First M87 event horizon telescope results. I. The Shadow of the supermassive black hole,
  Astrophys.\ J.\  {\bf 875} L1 (2019), 
  \href{https://arxiv.org/abs/1906.11238}{arXiv:1906.11238 [astro-ph.GA]}.

\bibitem{Akiyama:2019fyp} 
  K.~Akiyama {\it et al.} (Event Horizon Telescope Collaboration),
  First M87 event horizon telescope results. V. Physical origin of the asymmetric ring, 
  Astrophys.\ J.\  {\bf 875}, L5 (2019), 
  \href{https://arxiv.org/abs/1906.11242}{arXiv:1906.11242 [astro-ph.GA]}.
 
\bibitem{Grenzebach:2014fha} 
  A.~Grenzebach, V.~Perlick, and C.~Lämmerzahl,
 Photon regions and shadows of Kerr-Newman-NUT black holes with a cosmological constant,
  Phys.\ Rev.\ D {\bf 89},  124004 (2014), 
  \href{https://arxiv.org/abs/1403.5234}{arXiv:1403.5234 [gr-qc]}.

\bibitem{Li:2013jra} 
  Z.~Li and C.~Bambi,
  Measuring the Kerr spin parameter of regular black holes from their shadow, J.\ Cosmol.\ Astropart.\ Phys.\ {\bf 01}, 041 (2014), 
   \href{https://arxiv.org/abs/1309.1606}{arXiv:1309.1606 [gr-qc]}.
  
\bibitem{Abdujabbarov:2016hnw} 
  A.~Abdujabbarov, M.~Amir, B.~Ahmedov, and S.~G.~Ghosh,
  Shadow of rotating regular black holes,
  Phys.\ Rev.\ D {\bf 93}, 104004 (2016),
   \href{https://arxiv.org/abs/1604.03809}{arXiv:1604.03809 [gr-qc]}.
  
\bibitem{Dymnikova:2019vuz} 
  I.~Dymnikova and K.~Kraav,
  Identification of a regular black hole by its shadow,
  Universe {\bf 5},  163 (2019).
 
\bibitem{Virbhadra:2002ju}
K.~S.~Virbhadra and G.~F.~R.~Ellis, Gravitational lensing by naked singularities,
Phys. Rev. D \textbf{65}, 103004 (2002).

\bibitem{Virbhadra:2007kw}
K.~S.~Virbhadra and C.~R.~Keeton, 
Time delay and magnification centroid due to gravitational lensing by black holes and naked singularities,'
Phys. Rev. D \textbf{77}, 124014 (2008), 
\href{https://arxiv.org/abs/0710.2333}{arXiv:0710.2333 [gr-qc]}.

\bibitem{Zulianello:2020cmx}
A. Zulianello, R. Carballo-Rubio, S. Liberati, and S. Ansoldi, Electromagnetic tests of horizonless rotating black hole mimickers, Phys. Rev. D {\bf 103}, 064071 (2021), 
\href{https://arxiv.org/abs/2005.01837}{arXiv:2005.01837 [gr-qc]}. 
  
\bibitem{Zakharov2005} 
A. F. Zakharov, A. A. Nucita, F. DePaolis, and G. Ingrosso, 
Measuring the black hole parameters in the Galactic Center with RADIOASTRON,
New Astron. \textbf{10}, 479 (2005),
\href{https://arxiv.org/abs/astro-ph/0505286}{arXiv:0505286 [astro-ph]}.
%
\bibitem{Zakharov2005b}
A. F. Zakharov,  F. DePaolis, G.  Ingrosso, and A. A. Nucita,
Direct measurements of black hole charge  with future astrometrical missions,
Astron. Astrophys.  
\textbf{442}, 795 (2005), 
\href{https://arxiv.org/abs/astro-ph/0505286}{arXiv:0505286 [astro-ph]}.

\bibitem{Zakharov:2018awx} 
  A.~F.~Zakharov,
  Constraints on tidal charge of the supermassive black hole at the Galactic Center with trajectories of bright stars,
  Eur.\ Phys.\ J.\ C {\bf 78},  689 (2018),
  \href{https://arxiv.org/abs/1804.10374}{arXiv:1804.10374 [gr-qc]}.

\bibitem{Zajacek:2018ycb}
  M.~Zajaček, A.~Tursunov, A.~Eckart, and S.~Britzen,
On the charge of the Galactic Centre black hole,
  Mont.\ Not.\ R.\ Astron.\ Soc.\  {\bf 480}, 4408 (2018),
  \href{https://arxiv.org/abs/1808.07327}{arXiv:1808.07327 [astro-ph.GA]}.
  
\bibitem{Zajacek:2018vsj} 
  M.~Zajaček  et al.,
 Constraining the charge of the Galactic Centre black hole,
  J.\ Phys.\ Conf.\ Ser.\  {\bf 1258}, 012031 (2019),
\href{https://arxiv.org/abs/1812.03574}{arXiv:1812.03574 [astro-ph.GA]}.



\bibitem{Schroven:2018agz} 
  K.~Schroven, A.~Trova, E.~Hackmann and C.~L\"ammerzahl,
  Charged fluid structures around a rotating compact object with a magnetic dipole field,
  Phys.\ Rev.\ D {\bf 98}, 023017 (2018), 
  \href{https://arxiv.org/abs/1804.11286}{arXiv:1804.11286 [astro-ph.HE]}.

\bibitem{Trova:2018bsf} 
  A.~Trova, K.~Schroven, E.~Hackmann, V.~Karas,   Kov\'a\v{r}, and P.  Slan\'y,
  Equilibrium configurations of a charged fluid around a Kerr black hole,
  Phys. Rev. D {\bf 97},  104019 (2018),
  \href{https://arxiv.org/abs/1803.02262}{arXiv:1803.02262 [astro-ph.HE]}.
  
\bibitem{Bronnikov2011} K. A. Bronnikov and O. B. Zaslavskii, 
Neutral and charged matter in equilibrium with black holes, 
Phys. Rev. D  \textbf{84}, 084013 (2011), \href{https://arxiv.org/abs/1107.4701v3}{arXiv:1107.4701v3 [gr-qc]}.


\bibitem{Penrose1965} R. Penrose, Gravitational collapse and 
space-time
singularities, Phys. Rev. Lett. \textbf{14}, 57 (1965).

\bibitem{HawkingPenrose1970} S. W. Hawking and R. Penrose, 
The singularities of gravitational collapse and cosmology, Proc. R. Soc. London A \textbf{314}, 529 (1970).

\bibitem{HawkingEllis1973} S. W. Hawking and G. F. R. Ellis, {\em
The Large Scale Structure of Space and Time} (Cambridge University
Press, Cambridge 1973).

\bibitem{Penrose1978} R. Penrose, Singularities of spacetime,
in {\em Theoritical principles in astrophysics and relativity},
edited by. N. R. Lebovitz, W. H. Reid, and P. O. Vandervoort (Chicago University
Press, Chicago, 1978), p. 217.  

\bibitem{sakharov} A. D. Sakharov, The initial stage of an expanding
universe and appearance of a nonuniform distribution of matter,
Sov. Phys. JETP \textbf{22}, 241 (1966).

\bibitem{gliner1966} E. Gliner,
Algebraic properties of the energy-momentum
tensor and vacuum-like states of matter, Sov. Phys. JETP \textbf{22},
378 (1966).

\bibitem{dy92} I. G. Dymnikova, Vacuum nonsingular black hole,
Gen. Relativ. Gravit. \textbf{24}, 235 (1992).

\bibitem{dy00} I. G. Dymnikova, The algebraic structure of a cosmological term in spherically symmetric solutions,
Phys. Lett. B \textbf{472}, 33 (2000),
\href{https://arxiv.org/abs/gr-qc/9912116}{arXiv:gr-qc/9912116}.


\bibitem{dy03} I. Dymnikova, Spherically symmetric space-time with the regular de Sitter center, 
Int. J. Mod. Phys. D {\bf  12}, 1015  (2003),
\href{https://arxiv.org/abs/gr-qc/0304110}{arxiv:gr-qc/0304110 [gr-qc]}.

\bibitem{Bronnikov:2003yi}
K.~Bronnikov, A.~Dobosz, and I.~Dymnikova,
Nonsingular vacuum cosmologies with a variable cosmological term,
Classical Quantum  Gravity \textbf{20}, 3797 (2003),
\href{https://arxiv.org/abs/gr-qc/0302029}{arXiv:gr-qc/0302029 [gr-qc]}.

\bibitem{dy04} I. G. Dymnikova,
Regular electrically charged vacuum structures with de Sitter centre in nonlinear electrodynamics coupled to general relativity, 
Classical Quantum Gravity \textbf{21}, 4417 (2004),
\href{https://arxiv.org/abs/gr-qc/0407072v3}{ar arXiv:gr-qc/0407072v3}.

\bibitem{dy2011}
I.~Dymnikova, E.~Galaktionov, and A.~Poszwa,
Vacuum solitons with a de Sitter center as dark matter candidates,
Gravit. Cosmol. \textbf{17}, 38 (2011).


\bibitem{Lemos:2011vz} J. P. S. Lemos, V. T. Zanchin, Regular
black holes: Electrically charged solutions, Reissner-Nordström outside
a de Sitter core, Phys. Rev. D \textbf{83}, 124005 (2011), 
\href{https://arxiv.org/abs/1104.4790}{arXiv:1104.4790 [gr-qc]}.

\bibitem{Dymnikova:2015yma}
I.~Dymnikova and M.~Khlopov,
Regular black hole remnants and gravitons with de Sitter interior as heavy dark matter candidates probing inhomogeneity of early universe,
Int. J. Mod. Phys. D \textbf{24}, 1545002 (2015),
 \href{https://arxiv.org/abs/1510.01351}{arXiv:1510.01351 [gr-qc]}.


\bibitem{Khlopov:1985jw}
M.~Khlopov, B.~A.~Malomed, and I.~B.~Zeldovich,
Gravitational instability of scalar fields and formation of primordial black holes,
Mon. Not. R. Astron. Soc. \textbf{215}, 575-589 (1985).

\bibitem{Cotner:2019ykd}
E.~Cotner, A.~Kusenko, M.~Sasaki, and V.~Takhistov,
Analytic description of primordial black hole formation from scalar field fragmentation, 
J. Cosmol. Astropart. Phys. \textbf{10}, 077 (2019), 
\href{https://arxiv.org/abs/1907.10613}{arXiv:1907.10613 [astro-ph.CO]}.

\bibitem{Lyth:2005ze}
D.~H.~Lyth, K.~A.~Malik, M.~Sasaki, and I.~Zaballa,
Forming sub-horizon black holes at the end of inflation,
J. Cosmol. Astropart. Phys. \textbf{01}, 011 (2006), 
\href{https://arxiv.org/abs/astro-ph/0510647}{arXiv:astro-ph/0510647 [astro-ph]}.


\bibitem{br2006} K. A. Bronnikov and J. C. Fabris,
Regular phantom black holes, Phys. Rev. Lett. \textbf{96}, 251101 (2006),
\href{https://arxiv.org/abs/gr-qc/0511109}{arXiv:gr-qc/0511109}


 \bibitem{br2007-1} K. A. Bronnikov, H. Dehnen, and V. N. Melnikov,
 Regular black holes and black universes, 
 Gen. Relativ. Gravit. \textbf{39}, 973 (2007), \href{https://arxiv.org/abs/gr-qc/0611022}{arXiv:gr-qc/0611022}. 
 
 \bibitem{br2007-2} K. A. Bronnikov and I. Dymnikova, 
 Regular homogeneous T-models with vacuum dark fluid, 
 Classical Quantum Gravity \textbf{24}, 5803 (2007), \href{https://arxiv.org/abs/0705.2368}{arXiv:0705.2368 {[}gr-qc{]}}.
 
 \bibitem{ainou2011} M. Azreg-Aïnou, G. Clément, J. C. Fabris, and M. E. Rodrigues, 
 Phantom black holes and sigma models, Phys. Rev. D \textbf{83}, 124001 (2011), 
 \href{https://arxiv.org/abs/1102.4093}{arXiv:1102.4093 [hep-th]}.

\bibitem{bardeen1968} J. M. Bardeen, Non-singular general-relativistic gravitational collapse, 
in {\em Proceedings of GR5} (Tbilisi, URSS, 1968).


\bibitem{ab00} E. Ayón-Beato and A. García, The Bardeen model as a nonlinear magnetic monopole, Phys. Lett. B \textbf{493}, 149 (2000), 
\href{https://arxiv.org/abs/gr-qc/0009077}{arXiv:0009077 [gr-qc]}.

\bibitem{brcritic} K. A. Bronnikov, 
Comment on regular black hole in general relativity coupled to nonlinear electrodynamics, 
Phys. Rev. Lett. \textbf{85}, 4641 (2000).

\bibitem{ab98} E. Ayón-Beato and A. García,
Regular black hole in general relativity coupled to nonlinear electrodynamics,
Phys. Rev. Lett. \textbf{80}, 5056 (1998), 
\href{https://arxiv.org/abs/gr-qc/9911046}{arXiv:9911046 [gr-qc]}.


\bibitem{br01} K. A. Bronnikov,
Regular magnetic black holes and monopoles from nonlinear electrodynamics,
Phys. Rev. D \textbf{63}, 044005 (2001), 
\href{https://arxiv.org/abs/gr-qc/0006014}{arXiv:0006014 [gr-qc]}.

\bibitem{Hayward2006} S. A. Hayward, 
Formation and evaporation of nonsingular black holes,
Phys. Rev. Lett. \textbf{96}, 031103 (2006), 
\href{https://arxiv.org/abs/gr-qc/0506126}{arXiv:0506126 [gr-qc]}.


\bibitem{mat06} W. Berej, J. Matyjasek, D. Tryniecki, and M. Woronowicz,
Regular black holes in quadratic gravity, Gen. Relativ. Gravit. \textbf{38}, 885 (2006),
\href{https://arxiv.org/abs/hep-th/0606185}{arXiv:hep-th/0606185}.

\bibitem{mat08} J. Matyjasek, D. Tryniecki, and M. Klimek, 
Regular black holes in an asymptotically de Sitter universe, 
Mod. Phys. Lett. A \textbf{23}, 3377 (2008), \href{https://arxiv.org/abs/0809.2275}{arXiv:0809.2275[gr-qc]}.


\bibitem{br12} K. A. Bronnikov, R. A. Konoplya, and A. Zhidenko, 
Instabilities of wormholes and regular black holes supported by a phantom scalar field,
Phys. Rev. D {\bf 86}, 024028 (2012), \href{https://arxiv.org/abs/1205.2224v3}{ arXiv:1205.2224v3 [gr-qc]}.

\bibitem{fl13} A. Flachi and J. P. S. Lemos, 
Quasinormal modes of regular black holes, 
Phys. Rev. D {\bf 87}, 024034 (2013), \href{https://arxiv.org/abs/1211.6212v2}{ arXiv:1211.6212v2 [gr-qc]}.
  
\bibitem{macedo14} C. F. B. Macedo and L. C. B. Crispino, 
Absorption of planar massless scalar waves by Bardeen regular black holes,
Phys. Rev. D {\bf 90}, 064001 (2014); \href{https://arxiv.org/abs/1408.1779v2}{arXiv:1408.1779v2 [gr-qc]}.

\bibitem{balart14} L. Balart and E. C. Vagenas, 
Regular black holes with a nonlinear electrodynamics source, 
Phys. Rev. D {\bf 90}, 124045 (2014),
\href{https://arxiv.org/abs/1408.0306v2}{arXiv:1408.0306v2 [gr-qc]}.
 
 
\bibitem{Rodrigues:2015ayd} 
  M.~E.~Rodrigues, E.~L.~B.~Junior, G.~T.~Marques, and V.~T.~Zanchin, 
  Regular black holes in $f(R)$ gravity coupled to nonlinear electrodynamics,
   Phys.\ Rev.\ D {\bf 94},  024062 (2016) 
  [Addendum: Phys.\ Rev.\ D {\bf 94}, 049904 (2016)], \href{https://arxiv.org/abs/1511.00569}{arXiv:1511.00569 [gr-qc]}.
  
 \bibitem{br2017} K. A. Bronnikov, Nonlinear electrodynamics, regular black holes and wormholes,
  Int. J. Mod. Phys. D {\bf 27}, 1841005 (2018),
 \href{https://arxiv.org/abs/1711.00087}{arXiv:1711.00087 [gr-qc]}


\bibitem{anso2008} S. Ansoldi, 
Spherical black holes with regular center: A review of existing models including a recent realization with Gaussian sources,
\href{https://arxiv.org/abs/0802.0330}{arXiv:0802.0330 [gr-qc]}.

\bibitem{Nicolini:2008aj}
P.~Nicolini,
Noncommutative Black Holes, The final appeal to quantum gravity: A review,
Int. J. Mod. Phys. A \textbf{24}, 1229-1308 (2009),
 \href{https://arxiv.org/abs/1711.00087}{arXiv:0807.1939 [hep-th]}.

 \bibitem{Ali:2018boy}
M.~S.~Ali and S.~G.~Ghosh,
Exact $d$-dimensional Bardeen-de Sitter black holes and thermodynamics,
Phys. Rev. D \textbf{98}, 084025 (2018).


\bibitem{Nicolini:2019irw}
P.~Nicolini, E.~Spallucci, and M.~F.~Wondrak,
  Quantum corrected black holes from string T-duality,
  Phys.\ Lett.\ B {\bf 797}, 134888 (2019),
 \href{https://arxiv.org/abs/1902.11242}{arXiv:1902.11242 [gr-qc]}. 

\bibitem{Neves:2019ywx}
J.~C.~S.~Neves and A.~Saa,
Accretion of perfect fluids onto a class of regular black holes,
Ann. Phys. (Amsterdam) \textbf{420}, 168269 (2020).
\href{https://arxiv.org/abs/1906.03718}{arXiv:1906.03718 [gr-qc]}.


\bibitem{Easson:2020esh}
D.~A.~Easson, C.~Keeler, and T.~Manton,
Classical double copy of regular non-singular black holes,
\href{https://arxiv.org/abs/2007.16186}{arXiv:2007.16186 [gr-qc]}.


\bibitem{Simpson:2019mud}
A.~Simpson and M.~Visser,
Regular black holes with asymptotically Minkowski cores,
Universe \textbf{6}, 8 (2019), 
\href{https://arxiv.org/abs/1911.01020}{arXiv:1911.01020 [gr-qc]}.


\bibitem{Berry:2020ntz}
T.~Berry, A.~Simpson, and M.~Visser,
Photon spheres, ISCOs, and OSCOs: Astrophysical observables for regular black holes with asymptotically Minkowski cores, Universe {\bf 7}, 2 (2020),
\href{https://arxiv.org/abs/2008.13308}{arXiv:2008.13308 [gr-qc]}.


\bibitem{fro16} V. P. Frolov, 
Notes on nonsingular models of black holes, 
Phys. Rev. D {\bf 94}, 104056 (2016),
 \href{https://arxiv.org/abs/1609.01758v2}{arXiv:1609.01758v2[gr-qc]}.

\bibitem{fmm89} V. P. Frolov, M. A. Markov, and V. F. Mukhanov, 
Through a black hole into a new universe?,
Phys. Lett. B \textbf{216}, 272 (1989).

\bibitem{fmm90} V. P. Frolov, M. A. Markov, and V. F. Mukhanov,
Black holes as possible sources of closed and semiclosed worlds, Phys. Rev. D \textbf{41}, 383 (1990).


\bibitem{Israel1966} W. Israel, Singular hypersurfaces and
thin shell in general relativity, Nuovo Cimento B \textbf{44},
1 (1966); Erratum: Nouvo Cimento B  \textbf{48}, 463 (1966).

 
\bibitem{Boulware:1973tlq}
D.~G.~Boulware,
Naked Singularities, Thin shells, and the Reissner-Nordstr\"om metric,
Phys. Rev. D \textbf{8}, 2363 (1973).

\bibitem{Lake89} K. Lake and T. Zannias, 
Fitting de Sitter space into a black hole,
Phys. Lett. A {\bf 140}, 291 (1989).

\bibitem{Shanki:2003qm}
S.~Shankaranarayanan and N.~Dadhich,
Nonsingular black holes on the brane,
Int. J. Mod. Phys. D \textbf{13}, 1095 (2004),
\href{https://arxiv.org/abs/gr-qc/0306111}{ arXiv:gr-qc/0306111 [gr-qc]}.

\bibitem{uyf2012} N. Uchikata, S. Yoshida, and T. Futamase,
New solutions of charged regular black holes and their stability, 
Phys. Rev. D \textbf{86}, 084025 (2012),
\href{https://arxiv.org/abs/1209.3567}{arXiv:1209.3567 [gr-qc]}.

\bibitem{Lemos:2017vz} J. P. S. Lemos and V. T. Zanchin,
Regular black holes: Guilfoyle electrically charged solutions with a perfect fluid phantom core,
Phys. Rev. D \textbf{93}, 124012 (2016),
\href{https://arxiv.org/abs/1603.07359v2}{arXiv:1603.07359v2 [gr-qc]}.

\bibitem{Masa:2018elb} A.~D.~D.~Masa, E.~S.~de Oliveira, and V.~T.~Zanchin, 
New regular black hole solutions and other electrically charged compact objects with a de Sitter core and a matter layer,
 Int. J. Mod. Phys. D {\bf 27}, 1843015 (2018), \href{https://arxiv.org/abs/2008.04478}{arXiv:2008.04478 [gr-qc]}.


\bibitem{Halilsoy:2018kgy}
S.~Habib Mazharimousavi and M.~Halilsoy,
Regularization of the Reissner-Nordstr\"om black hole,
Eur. Phys. J. Plus \textbf{133}, 386 (2018),
\href{https://arxiv.org/abs/1703.05286}{arXiv:1703.05286 [physics.gen-ph]}.


\bibitem{bp90} R. Balbinot and E. Poisson, 
Stability of the Schwarzschild-de Sitter model, Phys. Rev. D \textbf{41}, 395 (1990).

\bibitem{coospercruz} F. I. Coosperstock and V. de la Cruz, 
Sources for the Reissner-Nordström metric, Gen. Relativ. Gravit. \textbf{9}, 835 (1978).

\bibitem{florides83} P. S. Florides, 
The complete field of charged perfect fluid spheres and of other static spherically symmetric charged distributions, 
J. Phys. A \textbf{16}, 1419 (1983).


\bibitem{Florides1977} P. S. Florides, 
The complete field of a general static spherically symmetric distribution of charge,
Nuovo Cimento A \textbf{42}, 343 (1977).

 
\bibitem{Chase1970}  J. E. Chase,
Gravitational instability and collapse of charged fluid shells, Nuovo Cimento B \textbf{67}, 136 (1970).

\bibitem{LemosZaslavskii2007} 
  J.~P.~S.~Lemos and O.~B.~Zaslavskii,
 Quasiblack holes: Definition and general properties,
  Phys.\ Rev.\ D {\bf 76}, 084030 (2007),
  \href{https://arxiv.org/abs/0707.1094}{arXiv:0707.1094 [gr-qc]}.
  
\bibitem{LemosZaslavskii2020}
J.~P.~S.~Lemos and O.~B.~Zaslavskii,
Compact objects in general relativity: From Buchdahl stars to quasiblack holes,
Int. J. Mod. Phys. D {\bf 29}, 2041019 (2020), 
 \href{https://arxiv.org/abs/2007.00665}{arXiv:2007.00665 [gr-qc]}.

 
\bibitem{Mazur2001} P. O. Mazur and E. Mottola,
Gravitational condensate stars: An alternative to black holes, 
\href{https://arxiv.org/abs/gr-qc/0109035v5}{ arXiv:gr-qc/0109035v5}.
 

\bibitem{Mazur:2004fk}
P.~O.~Mazur and E.~Mottola,
Gravitational vacuum condensate stars,
Proc. Natl. Acad. Sci. U.S.A. \textbf{101}, 9545 (2004),
\href{https://arxiv.org/abs/gr-qc/0407075} {arXiv:gr-qc/0407075 [gr-qc]}.


\bibitem{BNNCarter:2005pi}
B.~M.~N.~Carter,
Stable gravastars with generalised exteriors,
Classical Quantum Gravity \textbf{22}, 4551 (2005), 
\href{https://arxiv.org/abs/gr-qc/0509087} {arXiv:gr-qc/0509087 [gr-qc]}.


\bibitem{Horvat:2008ch}
D.~Horvat, S. Iliji\'c, and A. Marunovi\'c,
Electrically charged gravastar configurations,
Classical Quantum Gravity \textbf{26}, 025003 (2009),
 \href{https://arxiv.org/abs/0807.2051}{arXiv:0807.2051 [gr-qc]}.

\bibitem{Ghosh:2015ohi}
S.~Ghosh, F.~Rahaman, B.~K.~Guha, and S.~Ray,
Charged gravastars in higher dimensions,
Phys. Lett. B \textbf{767}, 380 (2017),
 \href{https://arxiv.org/abs/1511.05417}{arXiv:1511.05417 [physics.gen-ph]}.


\bibitem{Visser2004} M. Visser and D. L. Wiltshire, Stable gravastars--An alternative to black holes?,
Classical Quantum Gravity \textbf{21}, 1135 (2004), 
\href{https://arxiv.org/abs/gr-qc/0310107v2}{ arXiv:gr-qc/0310107v2}.
 
\bibitem{Lake1979}  K. Lake,  Thin spherical shells, Phys. Rev. D {\bf 19}, 2847 (1979).

\bibitem{Ishak:2001az}
M.~Ishak and K.~Lake,
Stability of transparent spherically symmetric thin shells and wormholes,
Phys. Rev. D \textbf{65}, 044011 (2002),
\href{https://arxiv.org/abs/gr-qc/0108058}{arXiv:gr-qc/0108058 [gr-qc]}.

\bibitem{Lobo:2005zu}
F. S. N. Lobo and P. Crawford, Stability analysis of dynamic thin shells, 
Classical Quantum Gravity \textbf{22}, 4869 (2005), 
\href{https://arxiv.org/abs/gr-qc/0507063}{arXiv:0507063 [gr-qc]}.

\bibitem{Dias:2010uh}
G.~A.~S.~Dias and J.~P.~S.~Lemos,
Thin-shell wormholes in $d$-dimensional general relativity: Solutions, properties, and stability,
Phys. Rev. D \textbf{82}, 084023 (2010), 
\href{https://arxiv.org/abs/1008.3376}{arXiv:1008.3376 [gr-qc]}.


\bibitem{Eiroa:2011nd}
E.~F.~Eiroa and C.~Simeone, Stability of charged thin shells,
Phys. Rev. D \textbf{83}, 104009 (2011),
\href{https://arxiv.org/abs/1102.1683} {arXiv:1102.1683 [gr-qc]}.


\bibitem{Varela2015} V. Varela, 
Note on linearized stability of Schwarzschild thin-shell wormholes with variable equations of state, 
Phys. Rev. D \textbf{92}, 044002 (2015), \href{https://arxiv.org/abs/1310.6420v5}{arXiv:1310.6420v5 [gr-qc]}.
 
\bibitem{Rosa:2020hex}
J.~L.~Rosa and P.~Pi\c{c}arra,
Existence and stability of relativistic fluid spheres supported by thin shells,
Phys. Rev. D \textbf{102}, 064009 (2020),
\href{https://arxiv.org/abs/2006.09854}{arXiv:2006.09854 [gr-qc]}.

\bibitem{Alestas:2020wwa}
G.~Alestas, G.~V.~Kraniotis and L.~Perivolaropoulos,
Existence and stability of static spherical fluid shells in a Schwarzschild-Rindler\textendash{}anti\textendash{}de Sitter metric,
Phys. Rev. D \textbf{102}, 104015 (2020),
\href{https://arxiv.org/abs/2005.11702}{arXiv:2005.11702 [gr-qc]}.

\bibitem{PerezBergliaffa:2020zzv}
S.~E.~Perez Bergliaffa, M.~Chiapparini, and L.~M.~Reyes,
Thermodynamical and dynamical stability of a self-gravitating uncharged thin shell,
Eur. Phys. J. C \textbf{80}, 719 (2020),
\href{https://arxiv.org/abs/2006.06766}{arXiv:2006.06766 [gr-qc]}.

\bibitem{Chirenti:2007mk}
C.~B.~M.~H.~Chirenti and L.~Rezzolla,
How to tell a gravastar from a black hole,
Classical Quantum Gravity \textbf{24}, 4191 (2007),
\href{https://arxiv.org/abs/0706.1513}{arXiv:0706.1513 [gr-qc]}.

\bibitem{Reviewer}  We would like to thank an anonymous reviewer for raising this question. 
See e.g. Ref.~\cite{Mann:2018jcf}) for a study on possible transitions of an initially timelike thin shell to a lightlike trajectory.

\bibitem{Mann:2018jcf} 
R.~B.~Mann, I.~Nagle, and D.~R.~Terno,
Transition to light-like trajectories in thin shell dynamics,
Nucl. Phys. B \textbf{936}, 19 (2018),
\href{https://arxiv.org/abs/1801.01981}{arXiv:1801.01981 [gr-qc]}.

\bibitem{Buchdahl} H. A. Buchdahl, General relativistic fluid spheres, Phys. Rev. \textbf{116}, 1027 (1959).

\bibitem{Andreasson09} H. Andr\'easson, Sharp bounds on the critical stability radius for relativistic charged spheres,
Commun. Math. Phys. \textbf{288}, 715 (2009), 
\href{https://arxiv.org/abs/0804.1882}{arXiv:0804.1882 [gr-qc]}.

\bibitem{Lemos:2015wfa}
J.~P.~S.~Lemos and V.~T.~Zanchin,
Sharp bounds on the radius of relativistic charged spheres: Guilfoyle's stars saturate the Buchdahl–Andréasson bound,
Classical Quantum Gravity \textbf{32}, 135009 (2015), 
 \href{https://arxiv.org/abs/1505.03863}{arXiv:1505.03863 [gr-qc]}.
\end{thebibliography}
\end{document}